\documentclass[journal=jpccck,manuscript=article]{achemso}

\usepackage[version=3]{mhchem} 
\usepackage{booktabs}

\author{Emin Aliyev}
\affiliation[Unknown University]
{UNAM-Institute of Materials Science and Nanotechnology, Bilkent University, Ankara 06800, Turkey}
\author{Arash Mobaraki}
\affiliation[Unknown University]
{UNAM-Institute of Materials Science and Nanotechnology, Bilkent University, Ankara 06800, Turkey}
\author{Hâldun Sevinçli}
\affiliation[Second University]{Department of Physics, Bilkent University, Ankara 06800, Turkey}
\author{Seymur Jahangirov}
\email{seymur@unam.bilkent.edu.tr}
\phone{+90 312 290 80 48}
\affiliation[Unknown University]
{UNAM-Institute of Materials Science and Nanotechnology, Bilkent University, Ankara 06800, Turkey}

\title[An \textsf{achemso} demo]
  {Thermodynamic favorability of the 1T phase over the 1H phase in group III metal monochalcogenide zigzag nanoribbons}

\keywords{Nanoribbons, Phase transition, Metal monochalcogenides, Electronic band structure}
\begin{document}

\begin{tocentry}
\includegraphics[width=8.25cm]{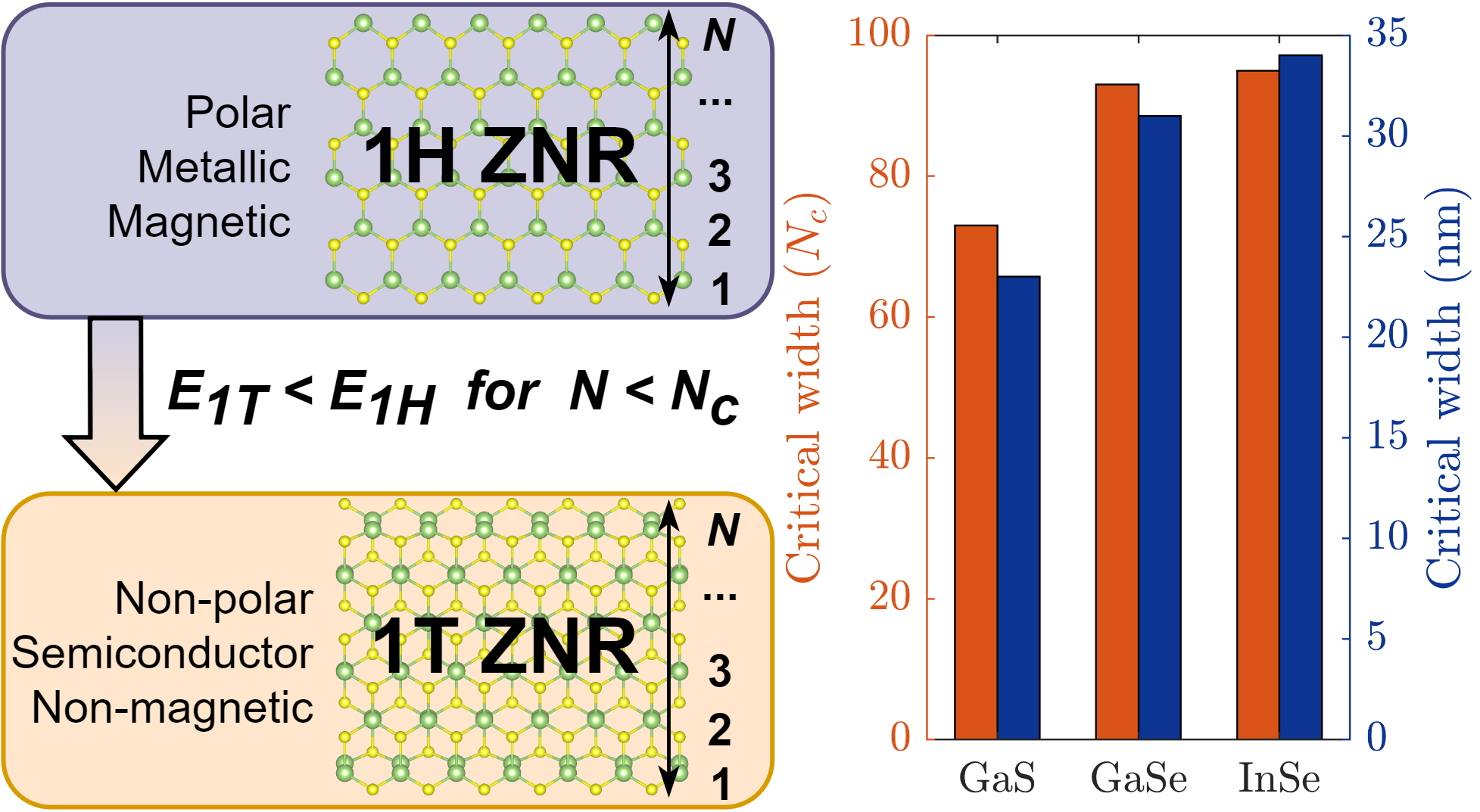}
\end{tocentry}

\begin{abstract}
Owing to the promising optoelectronic and thermoelectric properties of two-dimensional (2D) group III-VI materials (MXs), their nanoribbons (NRs) have attracted notable attention as an emerging class of quasi-one-dimensional (quasi-1D) nanostructures. Due to the fact that the most stable 2D monolayer polymorph of MXs is the 1H phase, to date, existing studies in the literature have predominantly focused on the NRs formed from 1H phase MXs. Nevertheless, NRs of the 1T phase have received little to no attention. Employing ab initio simulations based on density functional theory, we systematically compared the thermodynamic stability of hydrogen passivated and unpassivated 1T and 1H ZNRs of GaS, GaSe, and InSe. Our results reveal that non-polar 1T phase MX ZNRs are thermodynamically more favorable than polar 1H MX ZNRs at widths up to 34 nm, a range that is realizable through contemporary experimental fabrication techniques. Furthermore, unlike metallic 1H ZNRs, 1T ZNRs remain semiconductor and retain a Mexican-hat-shaped top valence bands. Complementarily, hydrogenation energies of 1T InSe ZNRs are positive, and due to the edge-localized states, the 1T unpassivated ZNRs possess nearly flat top valence bands. Our findings serve as a compass for subsequent synthesis pathways of group III-VI NRs.
\end{abstract}

\section{Introduction}
Since the isolation of 2D graphene monolayer from graphite~\cite{geim2007rise}, exploration of low-dimensional structures has attracted considerable attention of researchers. 
Among these structures, 2D group III metal monochalcogenides (MXs, M = Ga, In, and X = S, Se) are attracting heightened interest due to the remarkable variety of polymorphs with distinct physical properties~\cite{yu2024quasi,gutierrez2024unveiling, li2023emergence, hu2022prediction, wen2019two, bergeron2021polymorphism,grzonka2021novel,li2020various}. Unlike common 2D materials, they exhibit extraordinary nonparabolic inverted Mexican-hat-shaped band structure at the valence band maximum, which leads to flatter energy distribution, resulting in a van Hove singularity in the density of states~\cite{cai2019synthesis, ccinar2021ballistic}. The ground state of these MXs is hexagonal (H) phase ($D_{3h}$ point group), where a unit cell consists of metal atoms sandwiched between chalcogen atoms in X-M-M-X atomic arrangement. Few-layer and monolayer 2D structures of H-phase GaS, GaSe, and InSe were successfully synthesized with various methods~\cite{yang2019recent,yu2024review,bandurin2017high,brotons2016nanotexturing}. A recent study~\cite{jiang2023ballistic} reported fabrication of ballistic field effect transistor (FET) with mechanically exfoliated 2D InSe that outperformed any previously reported silicon FETs. In addition to FETs, studies report great potential of H-phase MXs in the fields of optoelectronics, spintronics, and thermoelectrics~\cite{yu2024review, xiong2023p}.

\begin{figure}[t!]  
  \centering  
  \includegraphics[width=6cm]{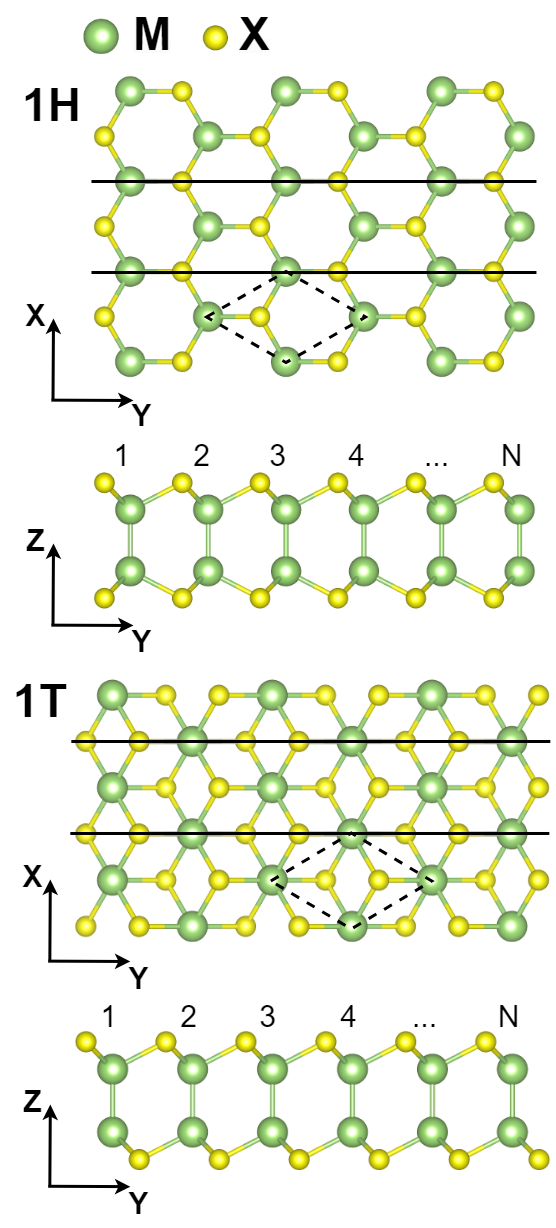}
  \caption{Schematic representations of 1T and 1H MX NRs and the unit cells of the corresponding 2D monolayers.}
  \label{fig:1}
  \end{figure}

Moreover, numerous theoretical investigations~\cite{li2020various,nitta2020first,zhou2016first} have shown that MXs can also crystallize in metastable staggered T phase ($D_{3d}$ point group), which is energetically quasi-degenerate phase,  as the energy difference between H and T phases is less than $k_{B}T$ at room temperature and can be traversed with strain or doping. The T phase structurally is similar to the H phase, with an exception of the atomic environment of the chalcogen atoms: in the X-M-M-X tetralayer the upper X atoms layer is rotated by 60° with respect to the lower one (see Fig.~\ref{fig:1})~\cite{zhou2016first}. In a recent study~\cite{gutierrez2024unveiling}, T-phase 2D GaS films were fabricated by chemical vapor deposition method. Furthermore, T-phase GaSe 2D layers were fabricated using molecular beam epitaxy method~\cite{yu2024quasi, grzonka2021novel}. In addition, 2D GaSe with T-phase-like layers formed due to intralayer sliding of Se atoms was reported to exhibit exotic nonvolatile memory behavior with a high channel current on/off ratio~\cite{li2023emergence}.

While 2D materials exhibit great potential and unique properties, by confining these 2D layers laterally ~\cite{cai2010atomically,chen2017fabrication}, one can obtain quasi-1D nanoribbons (NRs), which lead to the emergence of a broader spectrum of properties and enable fine-tuning of these properties by altering the width~\cite{son2006energy}, edge shapes, and various edge functionalizations of ribbons~\cite{bellunato2016chemistry,talirz2013termini}. Such adaptability and size made NRs attractive candidates for nanoscale applications in various fields, including spintronics~\cite{wang2024potential}, optoelectronics ~\cite{kumar2023electronic, he2023controlled}, catalysis ~\cite{aparna2023oxygen}, and biomedical applications ~\cite{hou2023pegylated}. Due to their intriguing properties and diverse applications, investigating the possible phase transition has been previously explored in NRs to expand the scope of their potential applications~\cite{li2021topological}. Such phase transitions are typically influenced by applying external electric and magnetic fields or imposing mechanical forces~\cite{huang2024strain,lu2018topological,
sivasubramani2018tunable,xie2023control,yang2022size,popple2023charge}. Furthermore, it has been demonstrated that the phase stability and electronic properties can be affected by the width of NRs~\cite{hong2023size,guller2015prediction,zan2018width,zan2022phase,zdetsis2023peculiar}.
 
The distinguishing properties of 2D MXs inspired many researchers to investigate their 1D counterparts as well. To date, various 1D MX nanostructures, such as nanobelts, nanowires, and NRs, have been fabricated and demonstrated to exhibit promising properties for the next generation of nanophotonics and optoelectronics~\cite{sutter2021optoelectronics, hauchecorne2021gallium, sutter2020vapor, wu2020controlled, xiong2016one, shen2009vapor, panda2008synthesis,arora2021recent}. Besides, 1H MX NRs were subjected to several theoretical studies that reported peculiar electronic and magnetic properties, such as intrinsic ferromagnetism (FM) in GaS NRs with zigzag edge (ZNRs)~\cite{wang2016electronic, zhou2014evidencing}. Notably, FM state in GaS ZNRs prevails even under tensile and compressive strains, with transition to AFM state being observed only at high compressive strains~\cite{wang2017strain}. Edge passivation with hydrogen also significantly affects the electronic properties of GaS NRs. Passivation of the GaS armchair-edge NRs results in an increase of band gap by 1 eV to 2.48 eV, which remains practically constant with increasing width~\cite{mosaferi2021band}. On the other hand, GaS and InSe ZNRs are reported to remain metallic even after passivation with hydrogen, except at very narrow widths ($N$ $\leq$ 4)~\cite{mosaferi2021band, yao2018electronic}. A different study reported comparable trends in the electronic properties of GaSe NRs, closely aligning with the observations in GaS NRs~\cite{zhou2015structures}. In addition, InSe ZNRs are also reported to be metallic and exhibit promising catalytic performance in hydrogen evolution reaction~\cite{wu2018modulation,cheng2019origin}.
 
 While all of the studies mentioned above have focused on the 1H MX NRs, the 1T MX NRs remain overlooked in the existing literature. In this paper, utilizing ab initio simulations based on density functional theory (DFT)~\cite{hohenberg1964inhomogeneous,kohn1965self}, we carried out a systematic analysis to explore the width-dependent variations in the energy difference between the 1T and 1H phases of MX ZNRs. Our investigation uncovers that in both cases of hydrogen-passivated ZNRs (P-ZNRs) and unpassivated ZNRs (UP-ZNRs), the 1T ZNRs are thermodynamically more favorable up to critical widths of about 23, 31, and 34 nm in the cases of GaS, GaSe, and InSe, respectively. Notably, 1T ZNRs are semiconductors with no built-in potential between their edges. 
 
\subsection{Computational Details}
All DFT calculations in this study were conducted using the Vienna Ab initio Software Package (VASP)~\cite{kresse1996efficient}. We employed projector augmented-wave pseudopotentials~\cite{kresse1999ultrasoft,blochl1994projector} within the generalized gradient approximation of Perdew, Burke, and Ernzerhof~\cite{perdew1996generalized,kresse1993ab,kresse1996efficiency}, with spin polarization considered. Ribbons are placed in the x-y plane (see Fig.~\ref{fig:1}), and a vacuum of 15 \text{\AA} is considered in non-periodic directions to eliminate interactions with images. 
The structures are fully relaxed using kinetic energy cutoff of 500 eV for the plane-wave basis and a $\mathrm{\Gamma}$-centered $10 \times 1 \times 1$ k-point mesh is used for sampling the Brillouin zone (BZ). The convergence tolerance for the total energy of the system is set to be less than $10^{-7}$ eV, and the Hellmann-Feynman forces are minimized to be less than $10^{-2}$ eV/\text{\AA}.

\subsection{Results and Discussion}

Defining $N$ as the number of the 2D monolayer unit cells forming the NR, we studied MX ZNRs ranging from  $N=4$ to $N=14$. As summarized in Table~S1, the calculated structural parameters and energy differences between the considered 2D MXs show perfect agreement with previous reports~\cite{zolyomi2014electrons, zhou2016first}. Besides, with increasing $N$, the lattice constants of the relaxed ZNRs converge toward the 2D monolayer lattice constants (see Fig.~S1). 

\begin{figure}[t!]  
  \centering  
  \includegraphics[width=14cm]{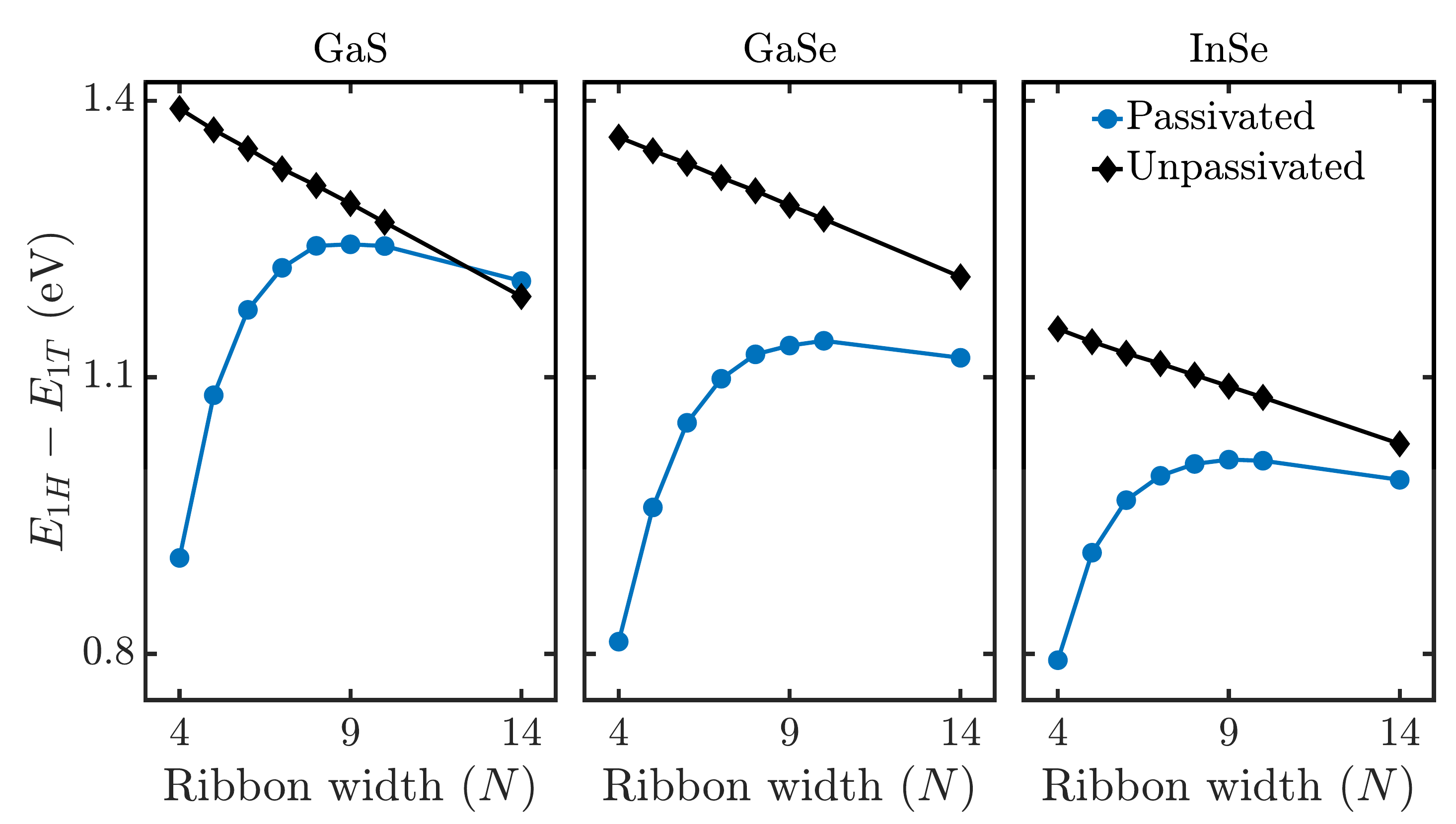}
  \caption{Energy difference between the 1H and 1T phases in passivated (blue circles) and unpassivated (black diamonds) cases with respect to $N$. Lines are guides for the eye.}
  \label{fig:2}
  \end{figure}
 Fig.~\ref{fig:2} illustrates the variation of the total ground state energy differences between the MX ZNRs of the 1H and 1T phases ($\Delta E = E_{1H}-E_{1T}$) as a function of $N$. The first noteworthy result deduced from this graph is that contrary to corresponding $2D$ monolayers, $\Delta E$ is positive in ZNRs of considered widths, indicating the thermodynamic favorability of the 1T phase. However, with increasing width, we observe a linear decrease in the $\Delta E$ of UP-ZNRs, while in the case of P-ZNRs, the $\Delta E$ starts to decrease after reaching a maximum value. Such behavior is expected since, at a certain critical width, the 1H should take over the 1T phase ($\Delta E < 0$). As described in ref~\cite{guller2015prediction}, this critical width can be determined by utilizing the formation energy of NRs. The formation energy for UP-NRs can be expressed as:
\begin{equation}\label{eq1}
E_{form}(N) = E_r(N) -N E_{2D}
\end{equation}  
where $E_{r} (N)$ is the total energy of the NR, $E_{2D}$ is the energy per formula unit of the corresponding 2D monolayer. For P-NRs, the formation energy of two hydrogen molecules is subtracted from the right side of Eq.~\ref{eq1}. Subsequently, the converged values of $E_{form}^{1H}$ and $E_{form}^{1T}$ with increasing $N$ can be used to determine $N_{c}$ as follows:
  
\begin{equation}\label{eq2}
N_{c} = \frac{E_{form}^{1H} - E_{form}^{1T}}{E_{2D}^{1T} - E_{2D}^{1H}}
\end{equation}
  
The formation energies as a function of $N$ are presented in Fig.~S2. One can notice the convergence. The $N=14$ values were used to calculate $E_{form}^{1H}$ and $E_{form}^{1T}$. The calculated $N_{c}$ values are given in Table~\ref{tbl:1}. The minimum $N_{c}$ is 72 (23 nm), which is found in the case of GaS P-ZNRs, and the maximum obtained value is 95 (34 nm), corresponding to InSe UP-ZNRs. Notably, the predicted widths are achievable by state-of-the-art fabrication methods~\cite{aslam2022single,hauchecorne2021gallium,panda2008synthesis,sutter2020vapor}, and significantly larger compared to TMD NRs, where the maximum reported critical width is 2.5 nm~\cite{guller2015prediction, zan2018width}.

\begin{table}[h]
\caption{ The calculated values of critical width in terms of $N_{c}$ (defined in Eq.~\ref{eq2}) where the crossover between the 1T and 1H phases of MX ZNRs occurs. The approximate width ($W$ in nm) is calculated using the lattice constant of the corresponding 1H 2D monolayer.}
\label{tbl:1}
\begin{tabular}{lcccc}
    \hline
    & \multicolumn{2}{c}{Unpassivated} & \multicolumn{2}{c}{~~Passivated} \\
    \cmidrule(r){2-3} \cmidrule(l){4-5}
    & $N_{c}$ & $W$ (\(\text{nm}\)) &$~~N_{c}$ & $W$ (\(\text{nm}\)) \\
    \hline
    GaS  & 73 & 23 &~~72 & 23 \\
    GaSe & 93 & 31 &~~87 & 29 \\
    InSe & 95 & 34 &~~92 & 33 \\
    \hline
\end{tabular}
\end{table}

 \begin{figure}[h!]  
  \centering  
  \includegraphics[width=13cm]{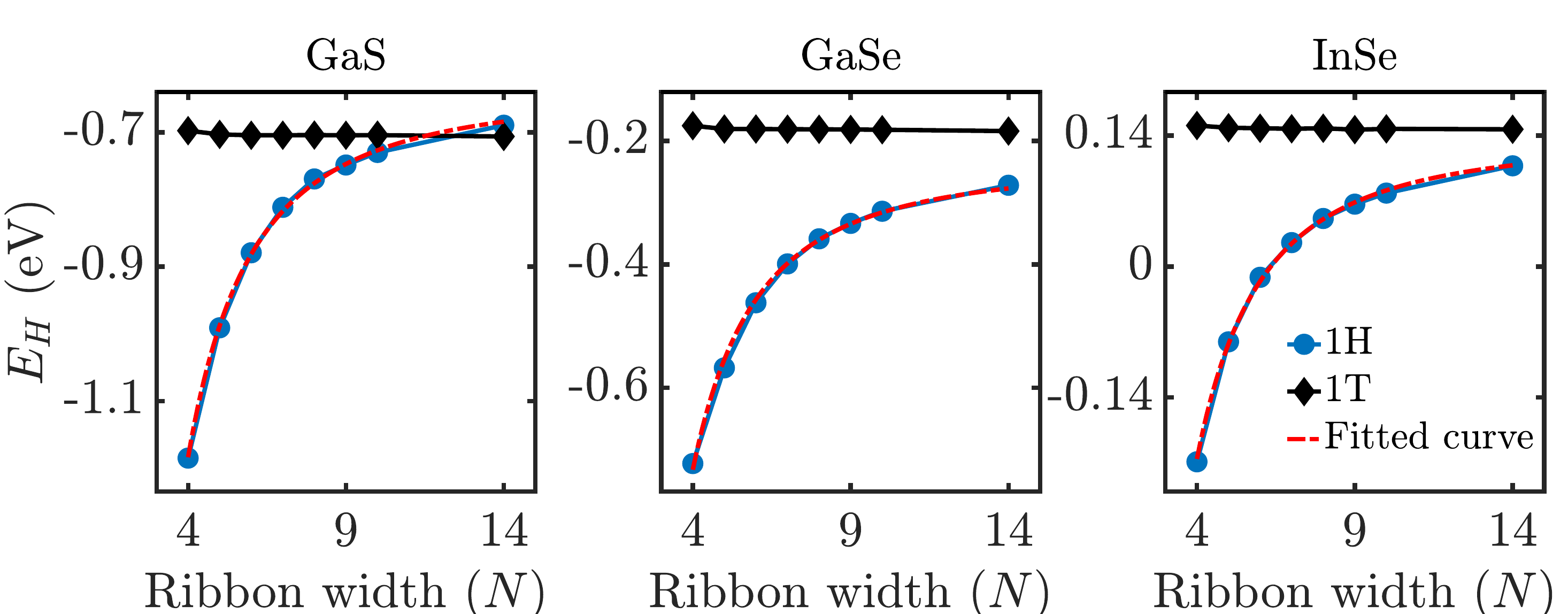}
  \caption{The calculated hydrogenation energies of 1T (black diamonds) and 1H (blue circles) MX ZNRs as a function of width. Solid lines are guides for the eye. The red dashed lines are $\frac{\alpha}{N^{2}}+\beta$ fits for describing the nonlinear behavior of 1H ZNRs.}
  \label{fig:3} 
  \end{figure}
  
  \begin{figure}[h!]  
  \centering  
  \includegraphics[width=14cm]{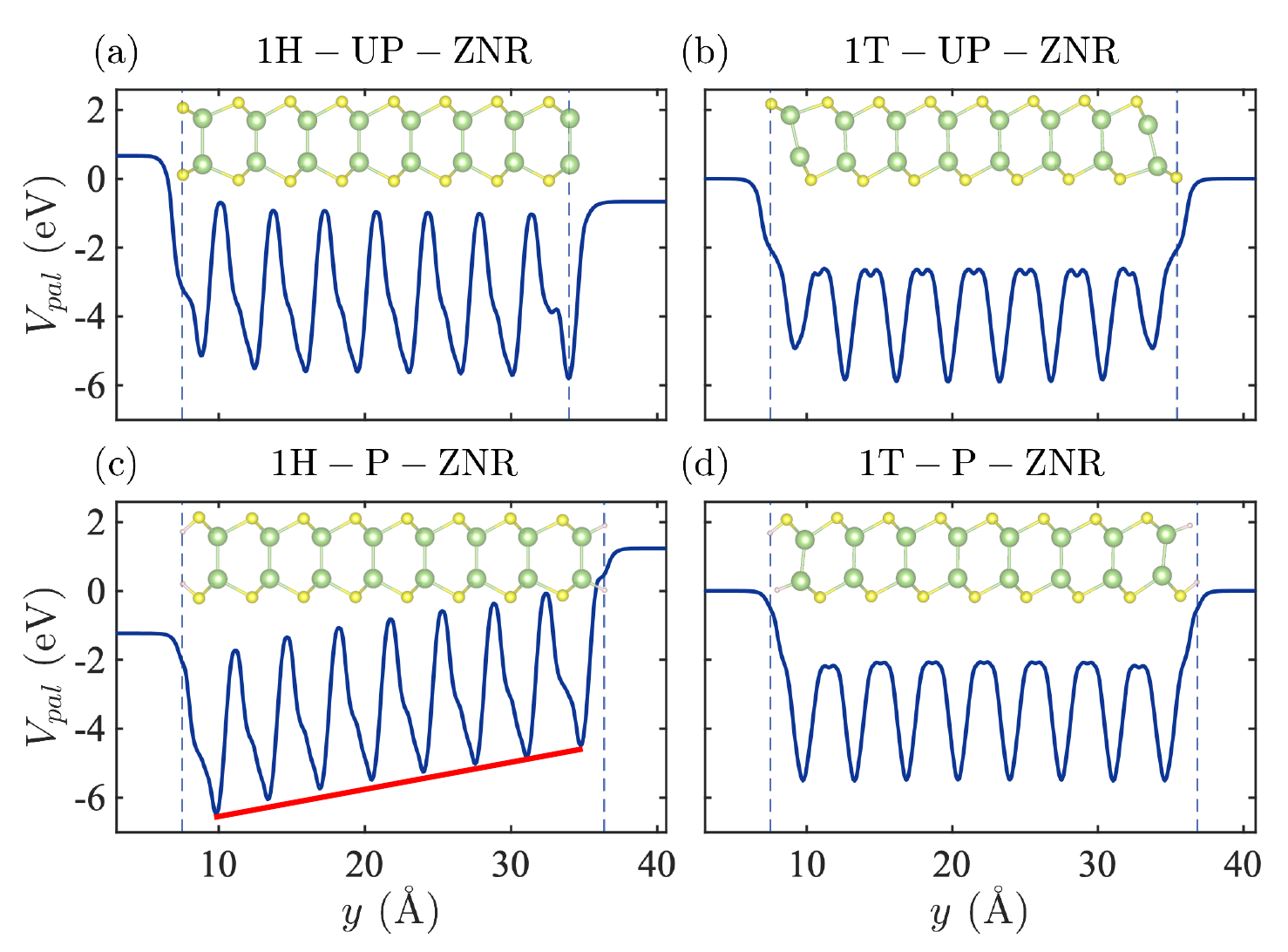}
  \caption{Planar average of the local electrostatic potential of (a) unpassivated 1H, (b) unpassivated 1T, (c) passivated 1H, and (d) passivated 1T $N=8$ InSe ZNRs. Vertical dashed lines show the position of the first and last atom along the width of the NRs.}
  \label{fig:4} 
  \end{figure}
The hydrogenation energies ($E_{H}$) depicted in Fig.~\ref{fig:3} give further insight into the nonlinear behavior of $\Delta E$ of passivated ZNRs. While the $E_{H}$ of 1T ZNRs remains practically constant, in the case of 1H ZNRs, the $\Delta E$ exhibits a nonlinear behavior that can be approximated by $\frac{\alpha}{N^{2}}+\beta$, as shown by the dashed red lines in Fig.~\ref{fig:3}. The origin of this nonlinearity in $E_{H}$ can be deduced from  Fig.~S2, where the $E_{form}$ of 1H-P-ZNRs exhibits a nonlinear behavior, while in the case of 1H-UP-ZNRs, it is practically constant. Another distinct feature of 1H-P-ZNRs can be seen in planar averaged local potentials ($V_{pal}$) presented in Figs.~\ref{fig:4},~S3, and~S4. Along the width of ZNRs, we observe oscillation in $V_{pal}$ with almost the same extrema except for the 1H-P-ZNRs, where there is practically a linear change of extrema, a behavior that was previously observed in the case of polar surfaces~\cite{song2008stabilizing,li2012ab} and NRs~\cite{yamanaka2017polarity}. An additional noteworthy feature that can be deduced from Fig.~\ref{fig:4} is the presence of a built-in electric field only in 1H ZNRs, which is a key factor in determining thermodynamic stability. As predicted theoretically by Zhong et al.~\cite{zhong2012prediction} and observed experimentally by Kuiper et al. ~\cite{kuiper2013control}, the stability of ultrathin films is determined by competition between built-in electric field and reconstruction processes, and as the number of layers decreases, the non-polar thin films become more favorable and compensate for the residual electrostatic energy. In a related context, we observe that the non-polar MX ZNRs become more stable when the ribbon widths are sufficiently small. Notably, a similar phenomenon has been observed in STEM images of few-layer 2D GaSe structures, where due to the presence of a built-in electric field, H to T phase transition occurred, which was utilized for ferroelectric switching~\cite{li2023emergence}.
  \begin{figure}[h!]  
  \centering  
  \includegraphics[width=12cm]{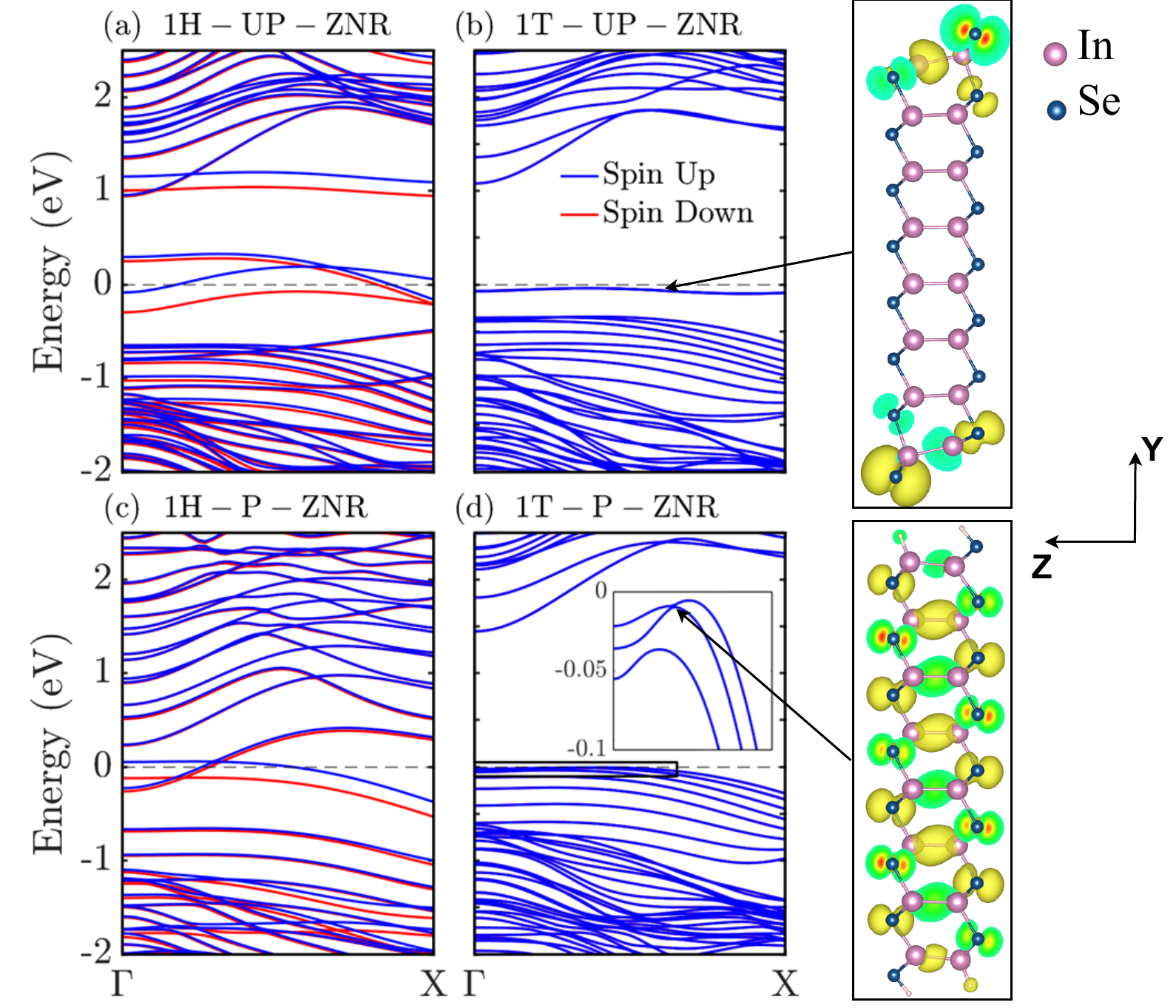}
  \caption{Spin-polarized band structures of (a) unpassivated 1H, (b) unpassivated 1T, (c) passivated 1H and (d) passivated 1T $N=8$ InSe ZNRs with magnified view of top valence bands in the inset. The callout figures depict the partial charge distribution summed over all k-points of the two highest valence bands.}
  \label{fig:5} 
  \end{figure}

The presence of built-in potential has prominent effects not only on the structural, but also on the electronic properties of NRs. It has been shown that applying an electric field in graphene and hexagonal boron nitride nanotubes and NRs leads to Wannier-Stark(WS)-like effect, which shifts and splits the degenerate electronic bands, reduces the band gap and results in semiconductor-to-metal transitions~\cite{park2008energy,filatov2024implications,alaei2013optical,wei2021electric,zhang2008energy,majhi2022enhanced}. Herein, we also observe that polar MX 1H ZNRs show metallic behavior even upon passivation, which is in agreement with previous studies~\cite{wu2018modulation,mosaferi2021band,wang2017strain, wang2016electronic,zhou2015structures,zhou2014evidencing}, while the non-polar MX 1T ZNRs, retain a band gap regardless of passivation and width, as presented in Figs.~\ref{fig:5}, S6, S7, S8, and S9. By comparing the electronic band structures of InSe 1H and 1T ZNR8s and their corresponding 2D monolayer supercells of equivalent size depicted in Fig. S10, one can notice the WS-like splitting of degenerate bands in the case of 1H-P-ZNR8, which possesses ladder-like $V_{pal}$ profile along the NR width (see Fig.~\ref{fig:4}(c)). In contrast, other NRs do not show such behavior and exhibit only slight breaking of the band degeneracy present in 2D due to atomic distortions as a result of confinement. Furthermore, the band gaps of 1T-P-ZNRs show a decreasing trend that converges to 2D monolayer band gaps as the width increases, while the band gaps of UP-ZNRs exhibits no substantial width dependence (see Fig.~S5). On top of it, similar to the 2D counterparts, 1T ZNRs possess a Mexican-hat-shaped top valence band. Furthermore, as shown in Figs.~\ref{fig:5} (d),~S6 (d), and ~S7 (d), 1T-UP-ZNRs exhibit a nearly flat, edge-localized top valence band (see Fig.~\ref{fig:5}). Notably, there is an increasing trend in ($E_{H}$) as the atomic number of the constituents in MX increases, reaching positive values in the case of InSe ZNRs (Fig.~\ref{fig:3}), indicating the possibility of 1T InSe ZNRs formation without hydrogen passivation. 

\subsection{Conclusion}

Employing DFT-based calculations, we thoroughly examined the width-dependent structural and electronic properties of hydrogen passivated and bare MX ZNRs. The remarkable finding of this study is the demonstration of the existing crossover between the 1T and 1H phases. Contrary to 2D monolayers, where 1H is the most thermodynamically favorable phase, in MX ZNRs, due to competition between
built-in electric field and reconstruction processes, the non-polar 1T phase becomes more favorable than the polar 1H phase below the critical widths spanning from 23 nm in GaS up to 34 nm in InSe ZNRs, a range achievable by state-of-the-art fabrication methods. The 1T-UP-ZNRs possess a nearly flat top valence band due to localized edge states, and band gaps that do not show substantial width-dependent variation. In 1T-P-ZNRs, the top valence band retains a Mexican-hat shape akin to their 2D monolayer counterparts, and band gaps exhibit a downward trend toward 2D monolayer values, creating a tunable platform for thermoelectric and optoelectronic applications. Another noteworthy observation is that, positive hydrogenation energies along with nearly flat top valence bands make InSe 1T-UP-ZNRs potential candidates for spintronic and thermoelectric applications. Our results lay the foundation for future experimental and theoretical investigations.

\begin{acknowledgement}

E.A., A.M. and S.J. acknowledge support from the TÜBİTAK project number 124F108. H.S. acknowledges support from the Air Force Office of Scientific Research (AFOSR, Award No. FA9550-21-1-0261). The calculations were performed at TÜBİTAK ULAKBİM, High Performance and Grid Computing Center (TRUBA resources) and National Center for High-Performance Computing of Turkey (UHeM) under grant number 1007742020.

\end{acknowledgement}

\begin{suppinfo}

Lattice constants of {MX} {2D} and {ZNR} structures, formation energies of {MX} {ZNR}s, bandgaps of 1{T} {MX} {ZNR}s, planar average of the local electrostatic potential of {GaS} and {GaS}e {ZNR}s for ${N} = 8$, spin-polarized band structures of {GaS} and {GaSe} {ZNR}s for ${N} = 8$; {InSe} 1{T-UP-ZNR}s and 1{T-P-ZNR}s for  ${N} = 4,5,6,7,9,10$; {InSe} 1{H} and 1{T} 2D $1 \times 8$ supercell and {ZNR}s for  ${N} = 8$.

\end{suppinfo}

\bibliography{achemso-demo}

\providecommand{\latin}[1]{#1}
\makeatletter
\providecommand{\doi}
  {\begingroup\let\do\@makeother\dospecials
  \catcode`\{=1 \catcode`\}=2 \doi@aux}
\providecommand{\doi@aux}[1]{\endgroup\texttt{#1}}
\makeatother
\providecommand*\mcitethebibliography{\thebibliography}
\csname @ifundefined\endcsname{endmcitethebibliography}  {\let\endmcitethebibliography\endthebibliography}{}
\begin{mcitethebibliography}{79}
\providecommand*\natexlab[1]{#1}
\providecommand*\mciteSetBstSublistMode[1]{}
\providecommand*\mciteSetBstMaxWidthForm[2]{}
\providecommand*\mciteBstWouldAddEndPuncttrue
  {\def\EndOfBibitem{\unskip.}}
\providecommand*\mciteBstWouldAddEndPunctfalse
  {\let\EndOfBibitem\relax}
\providecommand*\mciteSetBstMidEndSepPunct[3]{}
\providecommand*\mciteSetBstSublistLabelBeginEnd[3]{}
\providecommand*\EndOfBibitem{}
\mciteSetBstSublistMode{f}
\mciteSetBstMaxWidthForm{subitem}{(\alph{mcitesubitemcount})}
\mciteSetBstSublistLabelBeginEnd
  {\mcitemaxwidthsubitemform\space}
  {\relax}
  {\relax}

\bibitem[Geim and Novoselov(2007)Geim, and Novoselov]{geim2007rise}
Geim,~A.~K.; Novoselov,~K.~S. The rise of graphene. \emph{Nature materials} \textbf{2007}, \emph{6}, 183--191\relax
\mciteBstWouldAddEndPuncttrue
\mciteSetBstMidEndSepPunct{\mcitedefaultmidpunct}
{\mcitedefaultendpunct}{\mcitedefaultseppunct}\relax
\EndOfBibitem
\bibitem[Yu \latin{et~al.}(2024)Yu, Iddawela, Wang, Hilse, Thompson, Reifsnyder~Hickey, Sinnott, and Law]{yu2024quasi}
Yu,~M.; Iddawela,~S.~A.; Wang,~J.; Hilse,~M.; Thompson,~J.~L.; Reifsnyder~Hickey,~D.; Sinnott,~S.~B.; Law,~S. {Quasi}-{Van} der {Waals} {Epitaxial} {Growth} of $\gamma'$-{GaSe} {Nanometer}-{Thick} {Films} on {GaAs} (111) {B} {Substrates}. \emph{ACS nano} \textbf{2024}, \emph{18}, 17185--17196\relax
\mciteBstWouldAddEndPuncttrue
\mciteSetBstMidEndSepPunct{\mcitedefaultmidpunct}
{\mcitedefaultendpunct}{\mcitedefaultseppunct}\relax
\EndOfBibitem
\bibitem[Guti{\'e}rrez \latin{et~al.}(2024)Guti{\'e}rrez, Agresti, Juan, Dicorato, Giangregorio, Moreno, Garc{\'\i}a-Fern{\'a}ndez, Junquera, Armelao, and Losurdo]{gutierrez2024unveiling}
Guti{\'e}rrez,~Y.; Agresti,~F.; Juan,~D.; Dicorato,~S.; Giangregorio,~M.~M.; Moreno,~F.; Garc{\'\i}a-Fern{\'a}ndez,~P.; Junquera,~J.; Armelao,~L.; Losurdo,~M. {Unveiling} {Polymorphs} and {Polytypes} of the 2{D} {Layered} {Semiconducting} {Gallium} {Monosulfide}. \emph{Advanced Optical Materials} \textbf{2024}, \emph{12}, 2303002\relax
\mciteBstWouldAddEndPuncttrue
\mciteSetBstMidEndSepPunct{\mcitedefaultmidpunct}
{\mcitedefaultendpunct}{\mcitedefaultseppunct}\relax
\EndOfBibitem
\bibitem[Li \latin{et~al.}(2023)Li, Zhang, Yang, Zhou, Song, Cheng, Zhang, Feng, Wang, Lu, \latin{et~al.} others]{li2023emergence}
Li,~W.; Zhang,~X.; Yang,~J.; Zhou,~S.; Song,~C.; Cheng,~P.; Zhang,~Y.-Q.; Feng,~B.; Wang,~Z.; Lu,~Y. \latin{et~al.}  {Emergence} of ferroelectricity in a nonferroelectric monolayer. \emph{Nature Communications} \textbf{2023}, \emph{14}, 2757\relax
\mciteBstWouldAddEndPuncttrue
\mciteSetBstMidEndSepPunct{\mcitedefaultmidpunct}
{\mcitedefaultendpunct}{\mcitedefaultseppunct}\relax
\EndOfBibitem
\bibitem[Hu \latin{et~al.}(2022)Hu, Xu, Zhang, and Yu]{hu2022prediction}
Hu,~T.; Xu,~C.; Zhang,~A.; Yu,~P. {Prediction} of new phase 2{D} {C} 2h group {III} monochalcogenides with direct bandgaps and highly anisotropic carrier mobilities. \emph{Materials Advances} \textbf{2022}, \emph{3}, 2213--2221\relax
\mciteBstWouldAddEndPuncttrue
\mciteSetBstMidEndSepPunct{\mcitedefaultmidpunct}
{\mcitedefaultendpunct}{\mcitedefaultseppunct}\relax
\EndOfBibitem
\bibitem[Wen \latin{et~al.}(2019)Wen, Zhang, Guo, Shen, Sa, Lin, Zhou, and Sun]{wen2019two}
Wen,~C.; Zhang,~Z.; Guo,~Z.; Shen,~J.; Sa,~B.; Lin,~P.; Zhou,~J.; Sun,~Z. {Two}-dimensional {O}-phase group {III} monochalcogenides for photocatalytic water splitting. \emph{Journal of Physics: Condensed Matter} \textbf{2019}, \emph{32}, 065501\relax
\mciteBstWouldAddEndPuncttrue
\mciteSetBstMidEndSepPunct{\mcitedefaultmidpunct}
{\mcitedefaultendpunct}{\mcitedefaultseppunct}\relax
\EndOfBibitem
\bibitem[Bergeron \latin{et~al.}(2021)Bergeron, Lebedev, and Hersam]{bergeron2021polymorphism}
Bergeron,~H.; Lebedev,~D.; Hersam,~M.~C. {Polymorphism} in post-dichalcogenide two-dimensional materials. \emph{Chemical Reviews} \textbf{2021}, \emph{121}, 2713--2775\relax
\mciteBstWouldAddEndPuncttrue
\mciteSetBstMidEndSepPunct{\mcitedefaultmidpunct}
{\mcitedefaultendpunct}{\mcitedefaultseppunct}\relax
\EndOfBibitem
\bibitem[Grzonka \latin{et~al.}(2021)Grzonka, Claro, Molina-S{\'a}nchez, Sadewasser, and Ferreira]{grzonka2021novel}
Grzonka,~J.; Claro,~M.~S.; Molina-S{\'a}nchez,~A.; Sadewasser,~S.; Ferreira,~P.~J. {Novel} polymorph of {GaSe}. \emph{Advanced Functional Materials} \textbf{2021}, \emph{31}, 2104965\relax
\mciteBstWouldAddEndPuncttrue
\mciteSetBstMidEndSepPunct{\mcitedefaultmidpunct}
{\mcitedefaultendpunct}{\mcitedefaultseppunct}\relax
\EndOfBibitem
\bibitem[Li \latin{et~al.}(2020)Li, Li, and Wu]{li2020various}
Li,~X.; Li,~L.; Wu,~M. {Various} polymorphs of group {III}--{VI} ({GaSe}, {InSe}, {GaTe}) monolayers with quasi-degenerate energies: {Facile} phase transformations, high-strain plastic deformation, and ferroelastic switching. \emph{Materials Today Physics} \textbf{2020}, \emph{15}, 100229\relax
\mciteBstWouldAddEndPuncttrue
\mciteSetBstMidEndSepPunct{\mcitedefaultmidpunct}
{\mcitedefaultendpunct}{\mcitedefaultseppunct}\relax
\EndOfBibitem
\bibitem[Cai \latin{et~al.}(2019)Cai, Gu, Lin, Yu, Geohegan, and Xiao]{cai2019synthesis}
Cai,~H.; Gu,~Y.; Lin,~Y.-C.; Yu,~Y.; Geohegan,~D.~B.; Xiao,~K. {Synthesis} and emerging properties of 2{D} layered {III}--{VI} metal chalcogenides. \emph{Applied Physics Reviews} \textbf{2019}, \emph{6}\relax
\mciteBstWouldAddEndPuncttrue
\mciteSetBstMidEndSepPunct{\mcitedefaultmidpunct}
{\mcitedefaultendpunct}{\mcitedefaultseppunct}\relax
\EndOfBibitem
\bibitem[{\c{C}}{\i}nar \latin{et~al.}(2021){\c{C}}{\i}nar, Sarg{\i}n, Sevim, {\"O}zdamar, Kurt, and Sevin{\c{c}}li]{ccinar2021ballistic}
{\c{C}}{\i}nar,~M.~N.; Sarg{\i}n,~G.~{\"O}.; Sevim,~K.; {\"O}zdamar,~B.; Kurt,~G.; Sevin{\c{c}}li,~H. {Ballistic} thermoelectric transport properties of two-dimensional group {III}-{VI} monolayers. \emph{Physical Review B} \textbf{2021}, \emph{103}, 165422\relax
\mciteBstWouldAddEndPuncttrue
\mciteSetBstMidEndSepPunct{\mcitedefaultmidpunct}
{\mcitedefaultendpunct}{\mcitedefaultseppunct}\relax
\EndOfBibitem
\bibitem[Yang and Hao(2019)Yang, and Hao]{yang2019recent}
Yang,~Z.; Hao,~J. Recent progress in 2D layered {III--VI} semiconductors and their heterostructures for optoelectronic device applications. \emph{Advanced Materials Technologies} \textbf{2019}, \emph{4}, 1900108\relax
\mciteBstWouldAddEndPuncttrue
\mciteSetBstMidEndSepPunct{\mcitedefaultmidpunct}
{\mcitedefaultendpunct}{\mcitedefaultseppunct}\relax
\EndOfBibitem
\bibitem[Yu \latin{et~al.}(2024)Yu, Hilse, Zhang, Liu, Wang, and Law]{yu2024review}
Yu,~M.; Hilse,~M.; Zhang,~Q.; Liu,~Y.; Wang,~Z.; Law,~S. {Review} of {Nanolayered} {Post}-transition {Metal} {Monochalcogenides}: {Synthesis}, {Properties}, and {Applications}. \emph{ACS Applied Nano Materials} \textbf{2024}, \emph{7}, 28008--28026\relax
\mciteBstWouldAddEndPuncttrue
\mciteSetBstMidEndSepPunct{\mcitedefaultmidpunct}
{\mcitedefaultendpunct}{\mcitedefaultseppunct}\relax
\EndOfBibitem
\bibitem[Bandurin \latin{et~al.}(2017)Bandurin, Tyurnina, Yu, Mishchenko, Z\'{o}lyomi, Morozov, Kumar, Gorbachev, Kudrynskyi, Pezzini, \latin{et~al.} others]{bandurin2017high}
Bandurin,~D.~A.; Tyurnina,~A.~V.; Yu,~G.~L.; Mishchenko,~A.; Z\'{o}lyomi,~V.; Morozov,~S.~V.; Kumar,~R.~K.; Gorbachev,~R.~V.; Kudrynskyi,~Z.~R.; Pezzini,~S. \latin{et~al.}  High electron mobility, quantum Hall effect and anomalous optical response in atomically thin {InSe}. \emph{Nature nanotechnology} \textbf{2017}, \emph{12}, 223--227\relax
\mciteBstWouldAddEndPuncttrue
\mciteSetBstMidEndSepPunct{\mcitedefaultmidpunct}
{\mcitedefaultendpunct}{\mcitedefaultseppunct}\relax
\EndOfBibitem
\bibitem[Brotons-Gisbert \latin{et~al.}(2016)Brotons-Gisbert, Andres-Penares, Suh, Hidalgo, Abargues, Rodriguez-Canto, Segura, Cros, Tobias, Canadell, \latin{et~al.} others]{brotons2016nanotexturing}
Brotons-Gisbert,~M.; Andres-Penares,~D.; Suh,~J.; Hidalgo,~F.; Abargues,~R.; Rodriguez-Canto,~P.~J.; Segura,~A.; Cros,~A.; Tobias,~G.; Canadell,~E. \latin{et~al.}  Nanotexturing to enhance photoluminescent response of atomically thin indium selenide with highly tunable band gap. \emph{Nano letters} \textbf{2016}, \emph{16}, 3221--3229\relax
\mciteBstWouldAddEndPuncttrue
\mciteSetBstMidEndSepPunct{\mcitedefaultmidpunct}
{\mcitedefaultendpunct}{\mcitedefaultseppunct}\relax
\EndOfBibitem
\bibitem[Jiang \latin{et~al.}(2023)Jiang, Xu, Qiu, and Peng]{jiang2023ballistic}
Jiang,~J.; Xu,~L.; Qiu,~C.; Peng,~L.-M. {Ballistic} two-dimensional {InSe} transistors. \emph{Nature} \textbf{2023}, \emph{616}, 470--475\relax
\mciteBstWouldAddEndPuncttrue
\mciteSetBstMidEndSepPunct{\mcitedefaultmidpunct}
{\mcitedefaultendpunct}{\mcitedefaultseppunct}\relax
\EndOfBibitem
\bibitem[Xiong \latin{et~al.}(2023)Xiong, Xu, Feng, Zhang, Lin, and Chen]{xiong2023p}
Xiong,~Y.; Xu,~D.; Feng,~Y.; Zhang,~G.; Lin,~P.; Chen,~X. {P}-{Type} 2{D} {Semiconductors} for {Future} {Electronics}. \emph{Advanced Materials} \textbf{2023}, \emph{35}, 2206939\relax
\mciteBstWouldAddEndPuncttrue
\mciteSetBstMidEndSepPunct{\mcitedefaultmidpunct}
{\mcitedefaultendpunct}{\mcitedefaultseppunct}\relax
\EndOfBibitem
\bibitem[Nitta \latin{et~al.}(2020)Nitta, Yonezawa, Fleurence, Yamada-Takamura, and Ozaki]{nitta2020first}
Nitta,~H.; Yonezawa,~T.; Fleurence,~A.; Yamada-Takamura,~Y.; Ozaki,~T. {First}-principles study on the stability and electronic structure of monolayer {GaSe} with trigonal-antiprismatic structure. \emph{Physical Review B} \textbf{2020}, \emph{102}, 235407\relax
\mciteBstWouldAddEndPuncttrue
\mciteSetBstMidEndSepPunct{\mcitedefaultmidpunct}
{\mcitedefaultendpunct}{\mcitedefaultseppunct}\relax
\EndOfBibitem
\bibitem[Zhou and Zhuang(2016)Zhou, and Zhuang]{zhou2016first}
Zhou,~J.; Zhuang,~H.~L. {First}-{Principles} {Study} on the 1{T} {Phase} of {GaX} ({X}= {S}, {Se}) {Monolayers}. \emph{ChemistrySelect} \textbf{2016}, \emph{1}, 5779--5783\relax
\mciteBstWouldAddEndPuncttrue
\mciteSetBstMidEndSepPunct{\mcitedefaultmidpunct}
{\mcitedefaultendpunct}{\mcitedefaultseppunct}\relax
\EndOfBibitem
\bibitem[Cai \latin{et~al.}(2010)Cai, Ruffieux, Jaafar, Bieri, Braun, Blankenburg, Muoth, Seitsonen, Saleh, Feng, \latin{et~al.} others]{cai2010atomically}
Cai,~J.; Ruffieux,~P.; Jaafar,~R.; Bieri,~M.; Braun,~T.; Blankenburg,~S.; Muoth,~M.; Seitsonen,~A.~P.; Saleh,~M.; Feng,~X. \latin{et~al.}  {Atomically} precise bottom-up fabrication of graphene nanoribbons. \emph{Nature} \textbf{2010}, \emph{466}, 470--473\relax
\mciteBstWouldAddEndPuncttrue
\mciteSetBstMidEndSepPunct{\mcitedefaultmidpunct}
{\mcitedefaultendpunct}{\mcitedefaultseppunct}\relax
\EndOfBibitem
\bibitem[Chen \latin{et~al.}(2017)Chen, Cui, Ren, Zhang, Jin, Zhang, and Shih]{chen2017fabrication}
Chen,~Y.; Cui,~P.; Ren,~X.; Zhang,~C.; Jin,~C.; Zhang,~Z.; Shih,~C.-K. {Fabrication} of {MoSe2} nanoribbons via an unusual morphological phase transition. \emph{Nature communications} \textbf{2017}, \emph{8}, 15135\relax
\mciteBstWouldAddEndPuncttrue
\mciteSetBstMidEndSepPunct{\mcitedefaultmidpunct}
{\mcitedefaultendpunct}{\mcitedefaultseppunct}\relax
\EndOfBibitem
\bibitem[Son \latin{et~al.}(2006)Son, Cohen, and Louie]{son2006energy}
Son,~Y.-W.; Cohen,~M.~L.; Louie,~S.~G. {Energy} gaps in graphene nanoribbons. \emph{Physical review letters} \textbf{2006}, \emph{97}, 216803\relax
\mciteBstWouldAddEndPuncttrue
\mciteSetBstMidEndSepPunct{\mcitedefaultmidpunct}
{\mcitedefaultendpunct}{\mcitedefaultseppunct}\relax
\EndOfBibitem
\bibitem[Bellunato \latin{et~al.}(2016)Bellunato, Arjmandi~Tash, Cesa, and Schneider]{bellunato2016chemistry}
Bellunato,~A.; Arjmandi~Tash,~H.; Cesa,~Y.; Schneider,~G.~F. Chemistry at the {Edge} of {Graphene}. \emph{ChemPhysChem} \textbf{2016}, \emph{17}, 785--801\relax
\mciteBstWouldAddEndPuncttrue
\mciteSetBstMidEndSepPunct{\mcitedefaultmidpunct}
{\mcitedefaultendpunct}{\mcitedefaultseppunct}\relax
\EndOfBibitem
\bibitem[Talirz \latin{et~al.}(2013)Talirz, S\"{o}de, Cai, Ruffieux, Blankenburg, Jafaar, Berger, Feng, M{\"u}llen, Passerone, \latin{et~al.} others]{talirz2013termini}
Talirz,~L.; S\"{o}de,~H.; Cai,~J.; Ruffieux,~P.; Blankenburg,~S.; Jafaar,~R.; Berger,~R.; Feng,~X.; M{\"u}llen,~K.; Passerone,~D. \latin{et~al.}  {Termini} of bottom-up fabricated graphene nanoribbons. \emph{Journal of the American Chemical Society} \textbf{2013}, \emph{135}, 2060--2063\relax
\mciteBstWouldAddEndPuncttrue
\mciteSetBstMidEndSepPunct{\mcitedefaultmidpunct}
{\mcitedefaultendpunct}{\mcitedefaultseppunct}\relax
\EndOfBibitem
\bibitem[Wang \latin{et~al.}(2024)Wang, Fan, Yin, Li, Hu, Guo, Feng, Li, Zhang, Li, \latin{et~al.} others]{wang2024potential}
Wang,~Z.; Fan,~D.; Yin,~M.; Li,~H.; Hu,~H.; Guo,~F.; Feng,~Z.; Li,~J.; Zhang,~D.; Li,~Z. \latin{et~al.}  {Potential} applications of {C} 2 {N}-h2{D}/{BN} nanoribbon adsorption of transition metals in spintronic devices and magnetic storage devices. \emph{New Journal of Chemistry} \textbf{2024}, \emph{48}, 4699--4707\relax
\mciteBstWouldAddEndPuncttrue
\mciteSetBstMidEndSepPunct{\mcitedefaultmidpunct}
{\mcitedefaultendpunct}{\mcitedefaultseppunct}\relax
\EndOfBibitem
\bibitem[Kumar \latin{et~al.}(2023)Kumar, Pratap, Kumar, Mishra, Gwag, and Chakraborty]{kumar2023electronic}
Kumar,~S.; Pratap,~S.; Kumar,~V.; Mishra,~R.~K.; Gwag,~J.~S.; Chakraborty,~B. {Electronic}, transport, magnetic, and optical properties of graphene nanoribbons and their optical sensing applications: {A} comprehensive review. \emph{Luminescence} \textbf{2023}, \emph{38}, 909--953\relax
\mciteBstWouldAddEndPuncttrue
\mciteSetBstMidEndSepPunct{\mcitedefaultmidpunct}
{\mcitedefaultendpunct}{\mcitedefaultseppunct}\relax
\EndOfBibitem
\bibitem[He \latin{et~al.}(2023)He, Wang, Yan, Wan, Zhou, Zhang, Ye, Zhang, Shi, Jiang, \latin{et~al.} others]{he2023controlled}
He,~Z.; Wang,~K.; Yan,~C.; Wan,~L.; Zhou,~Q.; Zhang,~T.; Ye,~X.; Zhang,~Y.; Shi,~F.; Jiang,~S. \latin{et~al.}  {Controlled} preparation and device application of sub-5 nm graphene nanoribbons and graphene nanoribbon/carbon nanotube intramolecular heterostructures. \emph{ACS Applied Materials \& Interfaces} \textbf{2023}, \emph{15}, 7148--7156\relax
\mciteBstWouldAddEndPuncttrue
\mciteSetBstMidEndSepPunct{\mcitedefaultmidpunct}
{\mcitedefaultendpunct}{\mcitedefaultseppunct}\relax
\EndOfBibitem
\bibitem[Aparna and Chatanathodi(2023)Aparna, and Chatanathodi]{aparna2023oxygen}
Aparna,~M.; Chatanathodi,~R. {Oxygen} reduction and hydrogen evolution reactions on zigzag {ReS2} nanoribbons. \emph{Applied Surface Science} \textbf{2023}, \emph{618}, 156677\relax
\mciteBstWouldAddEndPuncttrue
\mciteSetBstMidEndSepPunct{\mcitedefaultmidpunct}
{\mcitedefaultendpunct}{\mcitedefaultseppunct}\relax
\EndOfBibitem
\bibitem[Hou \latin{et~al.}(2023)Hou, Cardo, Merino, Xu, Wetzl, Arnaiz, Luan, Mai, Criado, and Prato]{hou2023pegylated}
Hou,~H.; Cardo,~L.; Merino,~J.; Xu,~F.; Wetzl,~C.; Arnaiz,~B.; Luan,~X.; Mai,~Y.; Criado,~A.; Prato,~M. {PEGylated} bottom-up synthesized graphene nanoribbons loaded with camptothecin as potential drug carriers. \emph{Materials Today Chemistry} \textbf{2023}, \emph{33}, 101668\relax
\mciteBstWouldAddEndPuncttrue
\mciteSetBstMidEndSepPunct{\mcitedefaultmidpunct}
{\mcitedefaultendpunct}{\mcitedefaultseppunct}\relax
\EndOfBibitem
\bibitem[Li \latin{et~al.}(2021)Li, Sanz, Merino-D{\'\i}ez, Vilas-Varela, Garcia-Lekue, Corso, de~Oteyza, Frederiksen, Pe{\~n}a, and Pascual]{li2021topological}
Li,~J.; Sanz,~S.; Merino-D{\'\i}ez,~N.; Vilas-Varela,~M.; Garcia-Lekue,~A.; Corso,~M.; de~Oteyza,~D.~G.; Frederiksen,~T.; Pe{\~n}a,~D.; Pascual,~J.~I. {Topological} phase transition in chiral graphene nanoribbons: from edge bands to end states. \emph{Nature communications} \textbf{2021}, \emph{12}, 5538\relax
\mciteBstWouldAddEndPuncttrue
\mciteSetBstMidEndSepPunct{\mcitedefaultmidpunct}
{\mcitedefaultendpunct}{\mcitedefaultseppunct}\relax
\EndOfBibitem
\bibitem[Huang \latin{et~al.}(2024)Huang, Ke, Guan, Li, and Lou]{huang2024strain}
Huang,~A.; Ke,~S.; Guan,~J.-H.; Li,~J.; Lou,~W.-K. {Strain}-induced topological phase transition in graphene nanoribbons. \emph{Physical Review B} \textbf{2024}, \emph{109}, 045408\relax
\mciteBstWouldAddEndPuncttrue
\mciteSetBstMidEndSepPunct{\mcitedefaultmidpunct}
{\mcitedefaultendpunct}{\mcitedefaultseppunct}\relax
\EndOfBibitem
\bibitem[L{\"u} \latin{et~al.}(2018)L{\"u}, Xie, and Xie]{lu2018topological}
L{\"u},~X.~L.; Xie,~Y.; Xie,~H. {Topological} and magnetic phase transition in silicene-like zigzag nanoribbons. \emph{New Journal of Physics} \textbf{2018}, \emph{20}, 043054\relax
\mciteBstWouldAddEndPuncttrue
\mciteSetBstMidEndSepPunct{\mcitedefaultmidpunct}
{\mcitedefaultendpunct}{\mcitedefaultseppunct}\relax
\EndOfBibitem
\bibitem[Sivasubramani \latin{et~al.}(2018)Sivasubramani, Debroy, Acharyya, and Acharyya]{sivasubramani2018tunable}
Sivasubramani,~S.; Debroy,~S.; Acharyya,~S.~G.; Acharyya,~A. {Tunable} intrinsic magnetic phase transition in pristine single-layer graphene nanoribbons. \emph{Nanotechnology} \textbf{2018}, \emph{29}, 455701\relax
\mciteBstWouldAddEndPuncttrue
\mciteSetBstMidEndSepPunct{\mcitedefaultmidpunct}
{\mcitedefaultendpunct}{\mcitedefaultseppunct}\relax
\EndOfBibitem
\bibitem[Xie \latin{et~al.}(2023)Xie, Wang, Zhang, Wang, Yang, Li, and Li]{xie2023control}
Xie,~Y.; Wang,~B.; Zhang,~L.; Wang,~X.; Yang,~H.; Li,~G.; Li,~R.-W. {Control} of coexistent phase by rotation of magnetic field in a metamagnetic {FeRh} thin film. \emph{Journal of Magnetism and Magnetic Materials} \textbf{2023}, \emph{573}, 170674\relax
\mciteBstWouldAddEndPuncttrue
\mciteSetBstMidEndSepPunct{\mcitedefaultmidpunct}
{\mcitedefaultendpunct}{\mcitedefaultseppunct}\relax
\EndOfBibitem
\bibitem[Yang \latin{et~al.}(2022)Yang, Zong, Ding, and Sun]{yang2022size}
Yang,~Y.; Zong,~H.; Ding,~X.; Sun,~J. {Size}-dependent ferroic phase transformations in {GeSe} nanoribbons. \emph{Applied Physics Letters} \textbf{2022}, \emph{121}, 122903\relax
\mciteBstWouldAddEndPuncttrue
\mciteSetBstMidEndSepPunct{\mcitedefaultmidpunct}
{\mcitedefaultendpunct}{\mcitedefaultseppunct}\relax
\EndOfBibitem
\bibitem[Popple \latin{et~al.}(2023)Popple, Dogan, Hoang, Stonemeyer, Ercius, Bustillo, Cohen, and Zettl]{popple2023charge}
Popple,~D.; Dogan,~M.; Hoang,~T.~V.; Stonemeyer,~S.; Ercius,~P.; Bustillo,~K.~C.; Cohen,~M.; Zettl,~A. {Charge}-induced phase transition in encapsulated {Hf} {Te} 2 nanoribbons. \emph{Physical Review Materials} \textbf{2023}, \emph{7}, L013001\relax
\mciteBstWouldAddEndPuncttrue
\mciteSetBstMidEndSepPunct{\mcitedefaultmidpunct}
{\mcitedefaultendpunct}{\mcitedefaultseppunct}\relax
\EndOfBibitem
\bibitem[Hong \latin{et~al.}(2023)Hong, Deng, Ding, Sun, and Liu]{hong2023size}
Hong,~Y.; Deng,~J.; Ding,~X.; Sun,~J.; Liu,~J.~Z. {Size} {Limiting} {Elemental} {Ferroelectricity} in {Bi} {Nanoribbons}: {Observation}, {Mechanism}, and {Opportunity}. \emph{The Journal of Physical Chemistry Letters} \textbf{2023}, \emph{14}, 3160--3167\relax
\mciteBstWouldAddEndPuncttrue
\mciteSetBstMidEndSepPunct{\mcitedefaultmidpunct}
{\mcitedefaultendpunct}{\mcitedefaultseppunct}\relax
\EndOfBibitem
\bibitem[G{\"u}ller \latin{et~al.}(2015)G{\"u}ller, Llois, Goniakowski, and Noguera]{guller2015prediction}
G{\"u}ller,~F.; Llois,~A.~M.; Goniakowski,~J.; Noguera,~C. {Prediction} of structural and metal-to-semiconductor phase transitions in nanoscale {MoS} 2, {WS} 2, and other transition metal dichalcogenide zigzag ribbons. \emph{Physical Review B} \textbf{2015}, \emph{91}, 075407\relax
\mciteBstWouldAddEndPuncttrue
\mciteSetBstMidEndSepPunct{\mcitedefaultmidpunct}
{\mcitedefaultendpunct}{\mcitedefaultseppunct}\relax
\EndOfBibitem
\bibitem[Zan \latin{et~al.}(2018)Zan, Zhang, Yang, Yao, Li, and Yakobson]{zan2018width}
Zan,~W.; Zhang,~Z.; Yang,~Y.; Yao,~X.; Li,~S.; Yakobson,~B.~I. {Width}-dependent phase crossover in transition metal dichalcogenide nanoribbons. \emph{Nanotechnology} \textbf{2018}, \emph{30}, 075701\relax
\mciteBstWouldAddEndPuncttrue
\mciteSetBstMidEndSepPunct{\mcitedefaultmidpunct}
{\mcitedefaultendpunct}{\mcitedefaultseppunct}\relax
\EndOfBibitem
\bibitem[Zan \latin{et~al.}(2022)Zan, Huo, Mu, and Li]{zan2022phase}
Zan,~W.-Y.; Huo,~J.; Mu,~Y.-W.; Li,~S.-D. {Phase} crossover in transition metal dichalcogenide monolayers on metal substrates. \emph{Applied Surface Science} \textbf{2022}, \emph{599}, 153949\relax
\mciteBstWouldAddEndPuncttrue
\mciteSetBstMidEndSepPunct{\mcitedefaultmidpunct}
{\mcitedefaultendpunct}{\mcitedefaultseppunct}\relax
\EndOfBibitem
\bibitem[Zdetsis(2023)]{zdetsis2023peculiar}
Zdetsis,~A.~D. {Peculiar} electronic properties of wider armchair graphene nanoribbons: {Multiple} topological end-states and “phase transitions”. \emph{Carbon} \textbf{2023}, \emph{210}, 118042\relax
\mciteBstWouldAddEndPuncttrue
\mciteSetBstMidEndSepPunct{\mcitedefaultmidpunct}
{\mcitedefaultendpunct}{\mcitedefaultseppunct}\relax
\EndOfBibitem
\bibitem[Sutter \latin{et~al.}(2021)Sutter, French, Khosravi~Khorashad, Argyropoulos, and Sutter]{sutter2021optoelectronics}
Sutter,~P.; French,~J.~S.; Khosravi~Khorashad,~L.; Argyropoulos,~C.; Sutter,~E. {Optoelectronics} and nanophotonics of vapor--liquid--solid grown {GaSe} van der {Waals} nanoribbons. \emph{Nano letters} \textbf{2021}, \emph{21}, 4335--4342\relax
\mciteBstWouldAddEndPuncttrue
\mciteSetBstMidEndSepPunct{\mcitedefaultmidpunct}
{\mcitedefaultendpunct}{\mcitedefaultseppunct}\relax
\EndOfBibitem
\bibitem[Hauchecorne \latin{et~al.}(2021)Hauchecorne, Gity, Martin, Okuno, Bhattacharjee, Moeyaert, Rouchon, Hyot, Hurley, and Baron]{hauchecorne2021gallium}
Hauchecorne,~P.; Gity,~F.; Martin,~M.; Okuno,~H.; Bhattacharjee,~S.; Moeyaert,~J.; Rouchon,~D.; Hyot,~B.; Hurley,~P.~K.; Baron,~T. {Gallium} selenide nanoribbons on silicon substrates for photodetection. \emph{ACS Applied Nano Materials} \textbf{2021}, \emph{4}, 7820--7831\relax
\mciteBstWouldAddEndPuncttrue
\mciteSetBstMidEndSepPunct{\mcitedefaultmidpunct}
{\mcitedefaultendpunct}{\mcitedefaultseppunct}\relax
\EndOfBibitem
\bibitem[Sutter \latin{et~al.}(2020)Sutter, French, Sutter, Idrobo, and Sutter]{sutter2020vapor}
Sutter,~E.; French,~J.~S.; Sutter,~S.; Idrobo,~J.~C.; Sutter,~P. {Vapor}--liquid--solid growth and optoelectronics of gallium sulfide van der {Waals} nanowires. \emph{ACS nano} \textbf{2020}, \emph{14}, 6117--6126\relax
\mciteBstWouldAddEndPuncttrue
\mciteSetBstMidEndSepPunct{\mcitedefaultmidpunct}
{\mcitedefaultendpunct}{\mcitedefaultseppunct}\relax
\EndOfBibitem
\bibitem[Wu \latin{et~al.}(2020)Wu, Zhu, Wang, Kang, Xie, Wang, and Luo]{wu2020controlled}
Wu,~C.-Y.; Zhu,~H.; Wang,~M.; Kang,~J.; Xie,~C.; Wang,~L.; Luo,~L.-B. {Controlled} synthesis of {GaSe} microbelts for high-gain photodetectors induced by the electron trapping effect. \emph{Journal of Materials Chemistry C} \textbf{2020}, \emph{8}, 5375--5379\relax
\mciteBstWouldAddEndPuncttrue
\mciteSetBstMidEndSepPunct{\mcitedefaultmidpunct}
{\mcitedefaultendpunct}{\mcitedefaultseppunct}\relax
\EndOfBibitem
\bibitem[Xiong \latin{et~al.}(2016)Xiong, Zhang, Zhou, Jin, Li, and Zhai]{xiong2016one}
Xiong,~X.; Zhang,~Q.; Zhou,~X.; Jin,~B.; Li,~H.; Zhai,~T. {One}-step synthesis of p-type {GaSe} nanoribbons and their excellent performance in photodetectors and phototransistors. \emph{Journal of Materials Chemistry C} \textbf{2016}, \emph{4}, 7817--7823\relax
\mciteBstWouldAddEndPuncttrue
\mciteSetBstMidEndSepPunct{\mcitedefaultmidpunct}
{\mcitedefaultendpunct}{\mcitedefaultseppunct}\relax
\EndOfBibitem
\bibitem[Shen \latin{et~al.}(2009)Shen, Chen, Chen, and Zhou]{shen2009vapor}
Shen,~G.; Chen,~D.; Chen,~P.-C.; Zhou,~C. {Vapor}- solid growth of one-dimensional layer-structured gallium sulfide nanostructures. \emph{Acs Nano} \textbf{2009}, \emph{3}, 1115--1120\relax
\mciteBstWouldAddEndPuncttrue
\mciteSetBstMidEndSepPunct{\mcitedefaultmidpunct}
{\mcitedefaultendpunct}{\mcitedefaultseppunct}\relax
\EndOfBibitem
\bibitem[Panda \latin{et~al.}(2008)Panda, Datta, Sinha, Chaudhuri, Chavan, Patil, More, and Joag]{panda2008synthesis}
Panda,~S.~K.; Datta,~A.; Sinha,~G.; Chaudhuri,~S.; Chavan,~P.~G.; Patil,~S.~S.; More,~M.~A.; Joag,~D.~S. {Synthesis} of well-crystalline {GaS} nanobelts and their unique field emission behavior. \emph{The Journal of Physical Chemistry C} \textbf{2008}, \emph{112}, 6240--6244\relax
\mciteBstWouldAddEndPuncttrue
\mciteSetBstMidEndSepPunct{\mcitedefaultmidpunct}
{\mcitedefaultendpunct}{\mcitedefaultseppunct}\relax
\EndOfBibitem
\bibitem[Arora and Erbe(2021)Arora, and Erbe]{arora2021recent}
Arora,~H.; Erbe,~A. {Recent} progress in contact, mobility, and encapsulation engineering of {InSe} and {GaSe}. \emph{InfoMat} \textbf{2021}, \emph{3}, 662--693\relax
\mciteBstWouldAddEndPuncttrue
\mciteSetBstMidEndSepPunct{\mcitedefaultmidpunct}
{\mcitedefaultendpunct}{\mcitedefaultseppunct}\relax
\EndOfBibitem
\bibitem[Wang \latin{et~al.}(2016)Wang, Li, Zhang, Wang, and Ke]{wang2016electronic}
Wang,~B.-J.; Li,~X.-H.; Zhang,~L.-W.; Wang,~G.-D.; Ke,~S.-H. {Electronic} structures and edge effects of {Ga2S2} nanoribbons. \emph{Chinese Physics B} \textbf{2016}, \emph{25}, 107101\relax
\mciteBstWouldAddEndPuncttrue
\mciteSetBstMidEndSepPunct{\mcitedefaultmidpunct}
{\mcitedefaultendpunct}{\mcitedefaultseppunct}\relax
\EndOfBibitem
\bibitem[Zhou \latin{et~al.}(2014)Zhou, Li, Zhou, Zu, and Gao]{zhou2014evidencing}
Zhou,~Y.; Li,~S.; Zhou,~W.; Zu,~X.; Gao,~F. {Evidencing} the existence of intrinsic half-metallicity and ferromagnetism in zigzag gallium sulfide nanoribbons. \emph{Scientific reports} \textbf{2014}, \emph{4}, 5773\relax
\mciteBstWouldAddEndPuncttrue
\mciteSetBstMidEndSepPunct{\mcitedefaultmidpunct}
{\mcitedefaultendpunct}{\mcitedefaultseppunct}\relax
\EndOfBibitem
\bibitem[Wang \latin{et~al.}(2017)Wang, Li, Zhang, Wang, and Ke]{wang2017strain}
Wang,~B.-J.; Li,~X.-H.; Zhang,~L.-W.; Wang,~G.-D.; Ke,~S.-H. {Strain} engineering of electronic and magnetic properties of {Ga2S2} nanoribbons. \emph{Chinese Physics B} \textbf{2017}, \emph{26}, 057102\relax
\mciteBstWouldAddEndPuncttrue
\mciteSetBstMidEndSepPunct{\mcitedefaultmidpunct}
{\mcitedefaultendpunct}{\mcitedefaultseppunct}\relax
\EndOfBibitem
\bibitem[Mosaferi \latin{et~al.}(2021)Mosaferi, Sarsari, and Alaei]{mosaferi2021band}
Mosaferi,~M.; Sarsari,~I.~A.; Alaei,~M. {Band} structure engineering in gallium sulfide nanostructures. \emph{Applied Physics A} \textbf{2021}, \emph{127}, 123\relax
\mciteBstWouldAddEndPuncttrue
\mciteSetBstMidEndSepPunct{\mcitedefaultmidpunct}
{\mcitedefaultendpunct}{\mcitedefaultseppunct}\relax
\EndOfBibitem
\bibitem[Yao \latin{et~al.}(2018)Yao, Wang, Liu, and Sun]{yao2018electronic}
Yao,~A.-L.; Wang,~X.-F.; Liu,~Y.-S.; Sun,~Y.-N. Electronic Structure and I-V Characteristics of InSe Nanoribbons. \emph{Nanoscale Research Letters} \textbf{2018}, \emph{13}, 1--7\relax
\mciteBstWouldAddEndPuncttrue
\mciteSetBstMidEndSepPunct{\mcitedefaultmidpunct}
{\mcitedefaultendpunct}{\mcitedefaultseppunct}\relax
\EndOfBibitem
\bibitem[Zhou(2015)]{zhou2015structures}
Zhou,~J. {Structures} and electronic properties of {GaSe} and {GaS} nanoribbons. \emph{RSC advances} \textbf{2015}, \emph{5}, 94679--94684\relax
\mciteBstWouldAddEndPuncttrue
\mciteSetBstMidEndSepPunct{\mcitedefaultmidpunct}
{\mcitedefaultendpunct}{\mcitedefaultseppunct}\relax
\EndOfBibitem
\bibitem[Wu \latin{et~al.}(2018)Wu, Shi, Zhang, Ding, Wang, Cen, Guo, Pan, and Zhu]{wu2018modulation}
Wu,~M.; Shi,~J.-j.; Zhang,~M.; Ding,~Y.-m.; Wang,~H.; Cen,~Y.-l.; Guo,~W.-h.; Pan,~S.-h.; Zhu,~Y.-h. {Modulation} of electronic and magnetic properties in {InSe} nanoribbons: edge effect. \emph{Nanotechnology} \textbf{2018}, \emph{29}, 205708\relax
\mciteBstWouldAddEndPuncttrue
\mciteSetBstMidEndSepPunct{\mcitedefaultmidpunct}
{\mcitedefaultendpunct}{\mcitedefaultseppunct}\relax
\EndOfBibitem
\bibitem[Cheng \latin{et~al.}(2019)Cheng, Zhang, Guan, and Tao]{cheng2019origin}
Cheng,~X.; Zhang,~C.; Guan,~L.; Tao,~J. {Origin} of high hydrogen evolution activity on {InSe} nanoribbons: {A} first-principles study. \emph{International Journal of Hydrogen Energy} \textbf{2019}, \emph{44}, 24174--24183\relax
\mciteBstWouldAddEndPuncttrue
\mciteSetBstMidEndSepPunct{\mcitedefaultmidpunct}
{\mcitedefaultendpunct}{\mcitedefaultseppunct}\relax
\EndOfBibitem
\bibitem[Hohenberg and Kohn(1964)Hohenberg, and Kohn]{hohenberg1964inhomogeneous}
Hohenberg,~P.; Kohn,~W. {Inhomogeneous} electron gas. \emph{Physical review} \textbf{1964}, \emph{136}, B864\relax
\mciteBstWouldAddEndPuncttrue
\mciteSetBstMidEndSepPunct{\mcitedefaultmidpunct}
{\mcitedefaultendpunct}{\mcitedefaultseppunct}\relax
\EndOfBibitem
\bibitem[Kohn and Sham(1965)Kohn, and Sham]{kohn1965self}
Kohn,~W.; Sham,~L.~J. {Self}-consistent equations including exchange and correlation effects. \emph{Physical review} \textbf{1965}, \emph{140}, A1133\relax
\mciteBstWouldAddEndPuncttrue
\mciteSetBstMidEndSepPunct{\mcitedefaultmidpunct}
{\mcitedefaultendpunct}{\mcitedefaultseppunct}\relax
\EndOfBibitem
\bibitem[Kresse and Furthm{\"u}ller(1996)Kresse, and Furthm{\"u}ller]{kresse1996efficient}
Kresse,~G.; Furthm{\"u}ller,~J. {Efficient} iterative schemes for ab initio total-energy calculations using a plane-wave basis set. \emph{Physical review B} \textbf{1996}, \emph{54}, 11169\relax
\mciteBstWouldAddEndPuncttrue
\mciteSetBstMidEndSepPunct{\mcitedefaultmidpunct}
{\mcitedefaultendpunct}{\mcitedefaultseppunct}\relax
\EndOfBibitem
\bibitem[Kresse and Joubert(1999)Kresse, and Joubert]{kresse1999ultrasoft}
Kresse,~G.; Joubert,~D. {From} ultrasoft pseudopotentials to the projector augmented-wave method. \emph{Physical review B} \textbf{1999}, \emph{59}, 1758\relax
\mciteBstWouldAddEndPuncttrue
\mciteSetBstMidEndSepPunct{\mcitedefaultmidpunct}
{\mcitedefaultendpunct}{\mcitedefaultseppunct}\relax
\EndOfBibitem
\bibitem[Bl{\"o}chl(1994)]{blochl1994projector}
Bl{\"o}chl,~P.~E. {Projector} augmented-wave method. \emph{Physical review B} \textbf{1994}, \emph{50}, 17953\relax
\mciteBstWouldAddEndPuncttrue
\mciteSetBstMidEndSepPunct{\mcitedefaultmidpunct}
{\mcitedefaultendpunct}{\mcitedefaultseppunct}\relax
\EndOfBibitem
\bibitem[Perdew \latin{et~al.}(1996)Perdew, Burke, and Ernzerhof]{perdew1996generalized}
Perdew,~J.~P.; Burke,~K.; Ernzerhof,~M. {Generalized} gradient approximation made simple. \emph{Physical review letters} \textbf{1996}, \emph{77}, 3865\relax
\mciteBstWouldAddEndPuncttrue
\mciteSetBstMidEndSepPunct{\mcitedefaultmidpunct}
{\mcitedefaultendpunct}{\mcitedefaultseppunct}\relax
\EndOfBibitem
\bibitem[Kresse and Hafner(1993)Kresse, and Hafner]{kresse1993ab}
Kresse,~G.; Hafner,~J. {Ab} initio molecular dynamics for liquid metals. \emph{Physical review B} \textbf{1993}, \emph{47}, 558\relax
\mciteBstWouldAddEndPuncttrue
\mciteSetBstMidEndSepPunct{\mcitedefaultmidpunct}
{\mcitedefaultendpunct}{\mcitedefaultseppunct}\relax
\EndOfBibitem
\bibitem[Kresse and Furthm{\"u}ller(1996)Kresse, and Furthm{\"u}ller]{kresse1996efficiency}
Kresse,~G.; Furthm{\"u}ller,~J. {Efficiency} of ab-initio total energy calculations for metals and semiconductors using a plane-wave basis set. \emph{Computational materials science} \textbf{1996}, \emph{6}, 15--50\relax
\mciteBstWouldAddEndPuncttrue
\mciteSetBstMidEndSepPunct{\mcitedefaultmidpunct}
{\mcitedefaultendpunct}{\mcitedefaultseppunct}\relax
\EndOfBibitem
\bibitem[Z{\'o}lyomi \latin{et~al.}(2014)Z{\'o}lyomi, Drummond, and Fal'ko]{zolyomi2014electrons}
Z{\'o}lyomi,~V.; Drummond,~N.; Fal'ko,~V. {Electrons} and phonons in single layers of hexagonal indium chalcogenides from ab initio calculations. \emph{Physical Review B} \textbf{2014}, \emph{89}, 205416\relax
\mciteBstWouldAddEndPuncttrue
\mciteSetBstMidEndSepPunct{\mcitedefaultmidpunct}
{\mcitedefaultendpunct}{\mcitedefaultseppunct}\relax
\EndOfBibitem
\bibitem[Aslam \latin{et~al.}(2022)Aslam, Tran, Supina, Siri, Meunier, Watanabe, Taniguchi, Kralj, Teichert, Sheremet, \latin{et~al.} others]{aslam2022single}
Aslam,~M.~A.; Tran,~T.~H.; Supina,~A.; Siri,~O.; Meunier,~V.; Watanabe,~K.; Taniguchi,~T.; Kralj,~M.; Teichert,~C.; Sheremet,~E. \latin{et~al.}  {Single}-crystalline nanoribbon network field effect transistors from arbitrary two-dimensional materials. \emph{npj 2D Materials and Applications} \textbf{2022}, \emph{6}, 76\relax
\mciteBstWouldAddEndPuncttrue
\mciteSetBstMidEndSepPunct{\mcitedefaultmidpunct}
{\mcitedefaultendpunct}{\mcitedefaultseppunct}\relax
\EndOfBibitem
\bibitem[Song \latin{et~al.}(2008)Song, Akiyama, and Freeman]{song2008stabilizing}
Song,~J.-H.; Akiyama,~T.; Freeman,~A.~J. Stabilizing mechanism of the dipolar structure and its effects on formation of carriers in wurtzite $\{$0001$\}$ films: {InN} and {ZnO}. \emph{Physical Review B—Condensed Matter and Materials Physics} \textbf{2008}, \emph{77}, 035332\relax
\mciteBstWouldAddEndPuncttrue
\mciteSetBstMidEndSepPunct{\mcitedefaultmidpunct}
{\mcitedefaultendpunct}{\mcitedefaultseppunct}\relax
\EndOfBibitem
\bibitem[Li \latin{et~al.}(2012)Li, Gayles, Kioussis, Zhang, Grein, and Aqariden]{li2012ab}
Li,~J.; Gayles,~J.; Kioussis,~N.; Zhang,~Z.; Grein,~C.; Aqariden,~F. Ab initio studies of the unreconstructed polar {CdTe} (111) surface. \emph{Journal of electronic materials} \textbf{2012}, \emph{41}, 2745--2753\relax
\mciteBstWouldAddEndPuncttrue
\mciteSetBstMidEndSepPunct{\mcitedefaultmidpunct}
{\mcitedefaultendpunct}{\mcitedefaultseppunct}\relax
\EndOfBibitem
\bibitem[Yamanaka and Okada(2017)Yamanaka, and Okada]{yamanaka2017polarity}
Yamanaka,~A.; Okada,~S. Polarity control of {h-BN} nanoribbon edges by strain and edge termination. \emph{Physical Chemistry Chemical Physics} \textbf{2017}, \emph{19}, 9113--9117\relax
\mciteBstWouldAddEndPuncttrue
\mciteSetBstMidEndSepPunct{\mcitedefaultmidpunct}
{\mcitedefaultendpunct}{\mcitedefaultseppunct}\relax
\EndOfBibitem
\bibitem[Zhong \latin{et~al.}(2012)Zhong, Koster, and Kelly]{zhong2012prediction}
Zhong,~Z.; Koster,~G.; Kelly,~P.~J. Prediction of thickness limits of ideal polar ultrathin films. \emph{Physical Review B} \textbf{2012}, \emph{85}, 121411\relax
\mciteBstWouldAddEndPuncttrue
\mciteSetBstMidEndSepPunct{\mcitedefaultmidpunct}
{\mcitedefaultendpunct}{\mcitedefaultseppunct}\relax
\EndOfBibitem
\bibitem[Kuiper \latin{et~al.}(2013)Kuiper, Samal, Blank, ten Elshof, Rijnders, and Koster]{kuiper2013control}
Kuiper,~B.; Samal,~D.; Blank,~D.~H.; ten Elshof,~J.~E.; Rijnders,~G.; Koster,~G. Control of oxygen sublattice structure in ultra-thin SrCuO2 films studied by X-ray photoelectron diffraction. \emph{APL materials} \textbf{2013}, \emph{1}, 042113\relax
\mciteBstWouldAddEndPuncttrue
\mciteSetBstMidEndSepPunct{\mcitedefaultmidpunct}
{\mcitedefaultendpunct}{\mcitedefaultseppunct}\relax
\EndOfBibitem
\bibitem[Park and Louie(2008)Park, and Louie]{park2008energy}
Park,~C.-H.; Louie,~S.~G. Energy gaps and stark effect in boron nitride nanoribbons. \emph{Nano letters} \textbf{2008}, \emph{8}, 2200--2203\relax
\mciteBstWouldAddEndPuncttrue
\mciteSetBstMidEndSepPunct{\mcitedefaultmidpunct}
{\mcitedefaultendpunct}{\mcitedefaultseppunct}\relax
\EndOfBibitem
\bibitem[Filatov \latin{et~al.}(2024)Filatov, Pomogaeva, and Min]{filatov2024implications}
Filatov,~M.; Pomogaeva,~A.; Min,~S.~K. Implications of the edge states for the band structure of armchair graphene nanoribbons. \emph{Carbon Letters} \textbf{2024}, 1--13\relax
\mciteBstWouldAddEndPuncttrue
\mciteSetBstMidEndSepPunct{\mcitedefaultmidpunct}
{\mcitedefaultendpunct}{\mcitedefaultseppunct}\relax
\EndOfBibitem
\bibitem[Alaei and Sheikhi(2013)Alaei, and Sheikhi]{alaei2013optical}
Alaei,~R.; Sheikhi,~M. Optical absorption of graphene nanoribbon in transverse and modulated longitudinal electric field. \emph{Fullerenes, Nanotubes and Carbon Nanostructures} \textbf{2013}, \emph{21}, 183--197\relax
\mciteBstWouldAddEndPuncttrue
\mciteSetBstMidEndSepPunct{\mcitedefaultmidpunct}
{\mcitedefaultendpunct}{\mcitedefaultseppunct}\relax
\EndOfBibitem
\bibitem[Wei \latin{et~al.}(2021)Wei, Li, Jiang, and Cheng]{wei2021electric}
Wei,~Y.; Li,~W.; Jiang,~Y.; Cheng,~J. Electric field induced injection and shift currents in zigzag graphene nanoribbons. \emph{Physical Review B} \textbf{2021}, \emph{104}, 115402\relax
\mciteBstWouldAddEndPuncttrue
\mciteSetBstMidEndSepPunct{\mcitedefaultmidpunct}
{\mcitedefaultendpunct}{\mcitedefaultseppunct}\relax
\EndOfBibitem
\bibitem[Zhang and Guo(2008)Zhang, and Guo]{zhang2008energy}
Zhang,~Z.; Guo,~W. Energy-gap modulation of BN ribbons by transverse electric fields: First-principles calculations. \emph{Physical Review B—Condensed Matter and Materials Physics} \textbf{2008}, \emph{77}, 075403\relax
\mciteBstWouldAddEndPuncttrue
\mciteSetBstMidEndSepPunct{\mcitedefaultmidpunct}
{\mcitedefaultendpunct}{\mcitedefaultseppunct}\relax
\EndOfBibitem
\bibitem[Majhi \latin{et~al.}(2022)Majhi, Maiti, and Ganguly]{majhi2022enhanced}
Majhi,~J.; Maiti,~S.~K.; Ganguly,~S. Enhanced current rectification in graphene nanoribbons: effects of geometries and orientations of nanopores. \emph{Nanotechnology} \textbf{2022}, \emph{33}, 255704\relax
\mciteBstWouldAddEndPuncttrue
\mciteSetBstMidEndSepPunct{\mcitedefaultmidpunct}
{\mcitedefaultendpunct}{\mcitedefaultseppunct}\relax
\EndOfBibitem
\end{mcitethebibliography}
\clearpage



\section*{Supplementary Material}\label{sup}
\setcounter{figure}{0}
\setcounter{table}{0}
\setcounter{page}{1}
\renewcommand{\thepage}{S\arabic{page}}
\renewcommand{\thefigure}{S\arabic{figure}}
\renewcommand{\thetable}{S\arabic{table}}
\begin{table}[h!]
\centering

\caption{
The calculated energy differences ($\Delta E$) between 1T and 1H phase of 2D MXs and lattice constants of the structures are compared with previous works.
}
\resizebox{1\textwidth}{!}{%
\begin{tabular}{lccccccc}
\toprule
Material & \multicolumn{3}{c}{This Work} & \multicolumn{3}{c}{Previous Work} & Ref \\ 
\cmidrule(lr){2-4} \cmidrule(lr){5-7}
         & $\Delta E$ (meV) & 1H Lattice Constant (\text{\AA}) & 1T Lattice Constant (\text{\AA}) & $\Delta E$ (meV) & 1H Lattice Constant (\text{\AA}) & 1T Lattice Constant (\text{\AA}) \\ 
\midrule
GaS      & 20     & 3.63              & 3.64              & 20         & 3.63               & 3.64               & ~\cite{zhou2016first} \\ 
GaSe     & 15     & 3.82             & 3.83               & 15        & 3.82               & 3.83               & ~\cite{zhou2016first} \\ 
InSe     & 13    & 4.09             & 4.09              & 13       & 4.09               & 4.09               & ~\cite{zolyomi2014electrons} \\ 
\bottomrule
\end{tabular}
}
\label{tab:s1}
\end{table}


\begin{figure}[h!]
\centering
\includegraphics[width=13cm]{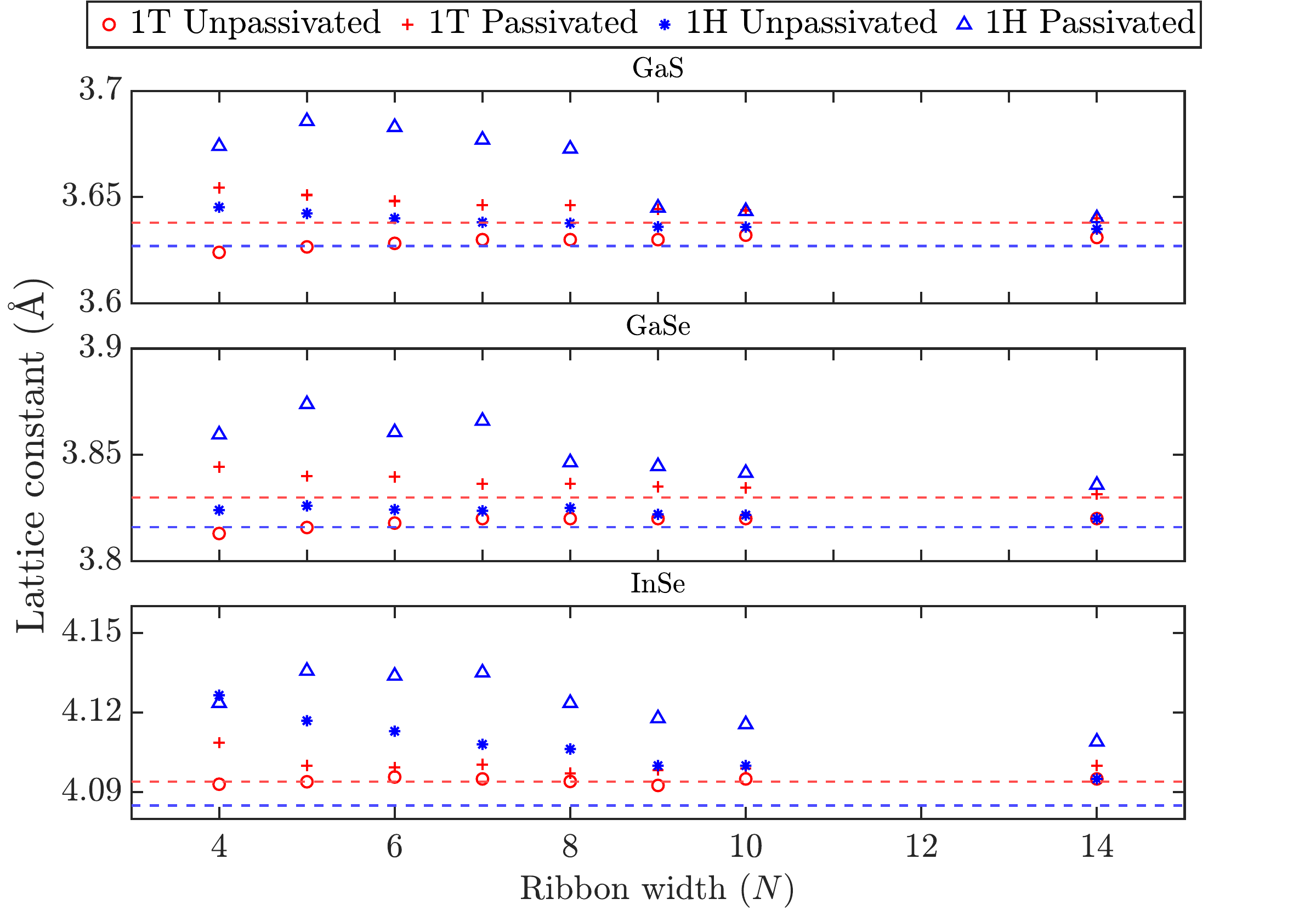}
\caption{Variation of the calculated lattice constants of MX ribbons with ribbon width. Red and blue dashed lines show the calculated lattice constants of 2D 1T and 1H MXs, respectively.}
\label{fig:supp1}
\end{figure}


\begin{figure}[h!]
\centering
\includegraphics[width=12cm]{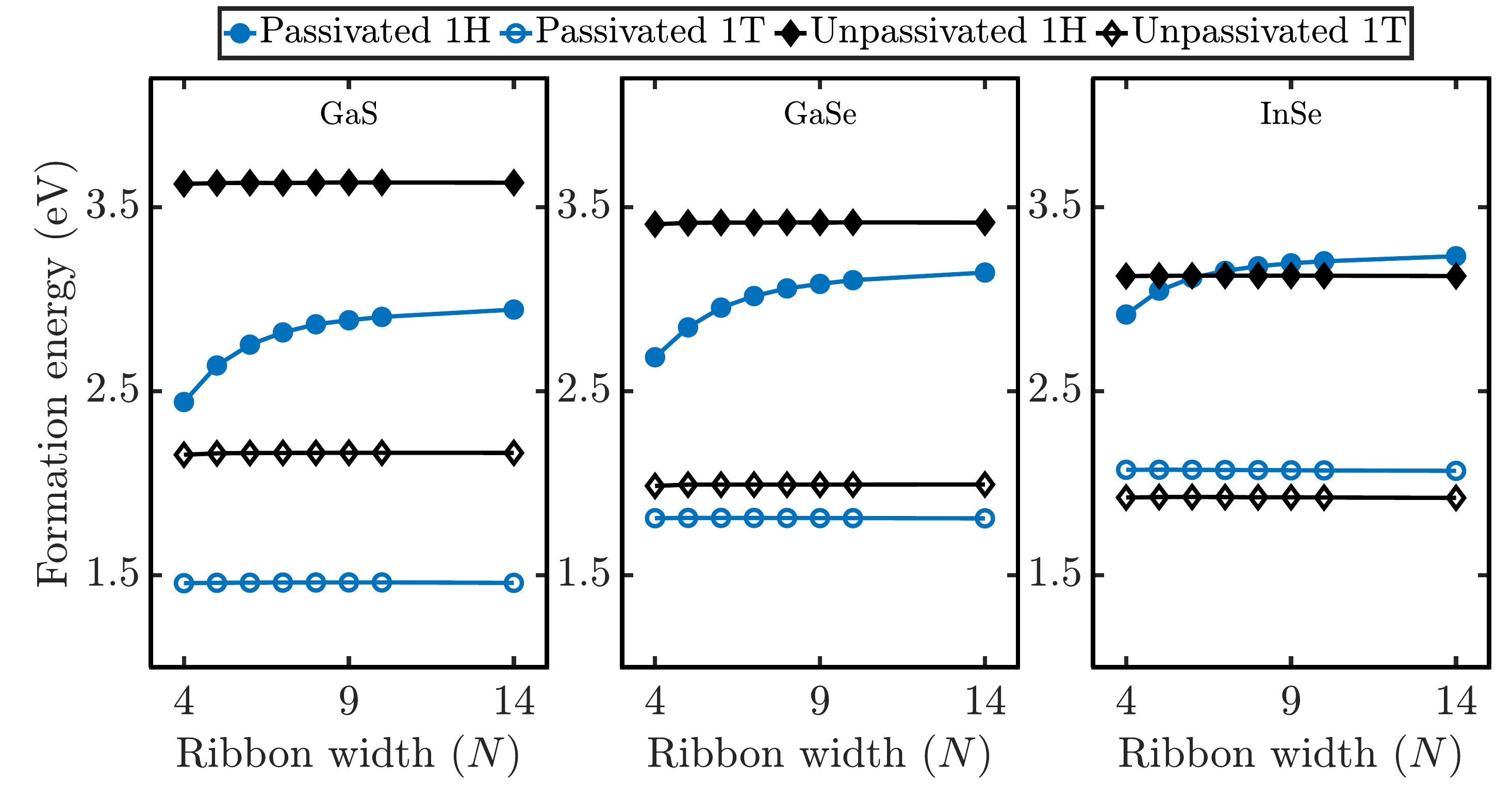}
\caption{Calculated formation energies of MX NRs plotted against ribbon width. Solid Lines are guides to the eye.}
\label{fig:supp2}
\end{figure}

\begin{figure}[h!]
\centering
\includegraphics[width=0.7\linewidth]{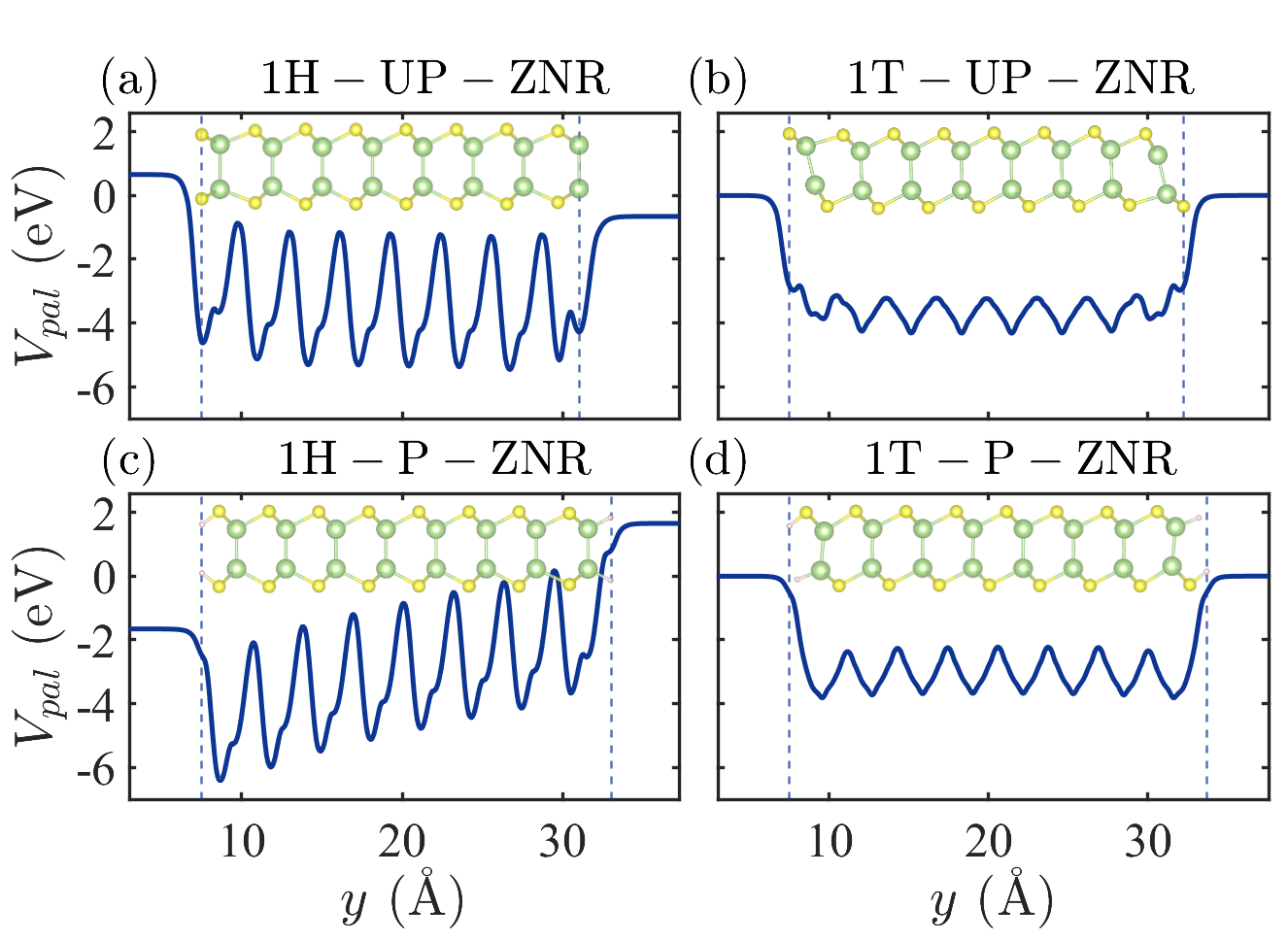}
\caption{Planar average of the local electrostatic potential of (a) unpassivated 1H, (b) unpassivated 1T, (c) passivated 1H, and (d) passivated 1T $N=8$ GaS ZNRs. Vertical dashed lines show the position of the first and last atom along the width of the NRs.}
\label{fig:supp3}
\end{figure}


\begin{figure}[h!]
\includegraphics[width=0.7\linewidth]{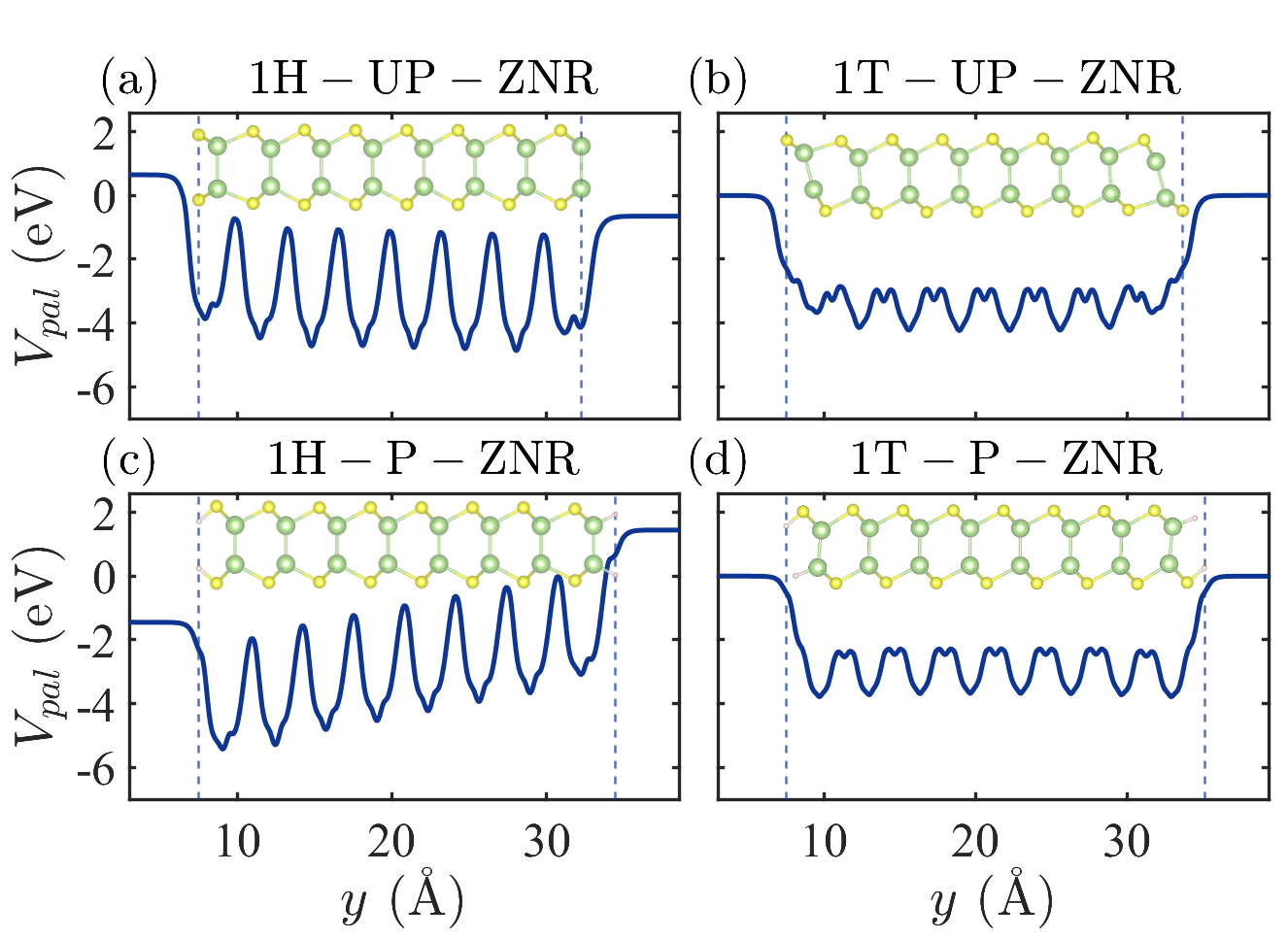}
\caption{Planar average of the local electrostatic potential of (a) unpassivated 1H, (b) unpassivated 1T, (c) passivated 1H, and (d) passivated 1T $N=8$ GaSe ZNRs. Vertical dashed lines show the position of the first and last atom along the width of the NRs.}
\label{fig:supp4}
\end{figure}

\begin{figure}[h!]
\centering
\includegraphics[width=0.9\linewidth]{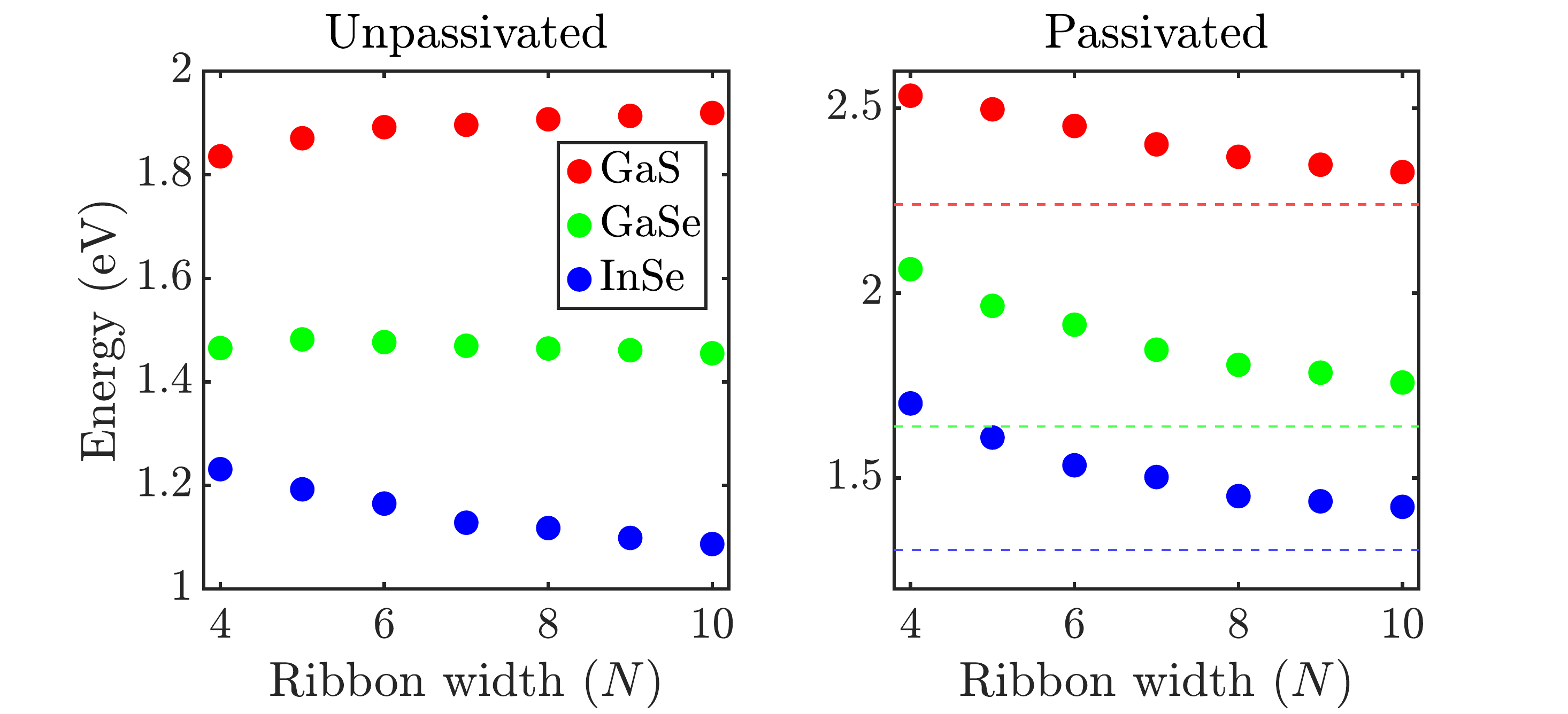}
\caption{Variation of the bandgap of 1T MX ZNRs with ribbon width. Dashed lines represent the band gaps of 2D MXs, and the symbols represent the 1T ZNRs. Colors are consistent to indicate the respective materials.}
\label{fig:supp5}
\end{figure}

\begin{figure}[h!]
\centering
\includegraphics[width=7.8cm]{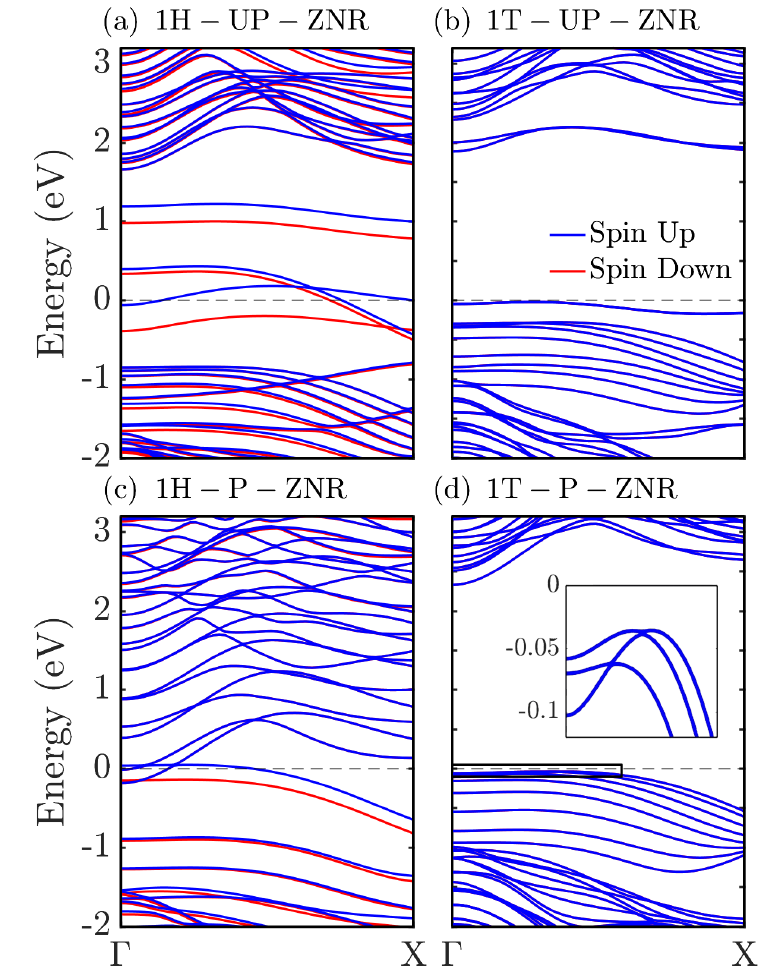}
\caption{Spin-polarized band structures of (a) unpassivated 1H, (b) unpassivated 1T, (c) passivated 1H and (d) passivated 1T $N=8$ GaS ZNRs.}
\label{fig:supp6}
\end{figure}

\begin{figure}[h!]
\centering
\includegraphics[width=7.8cm]{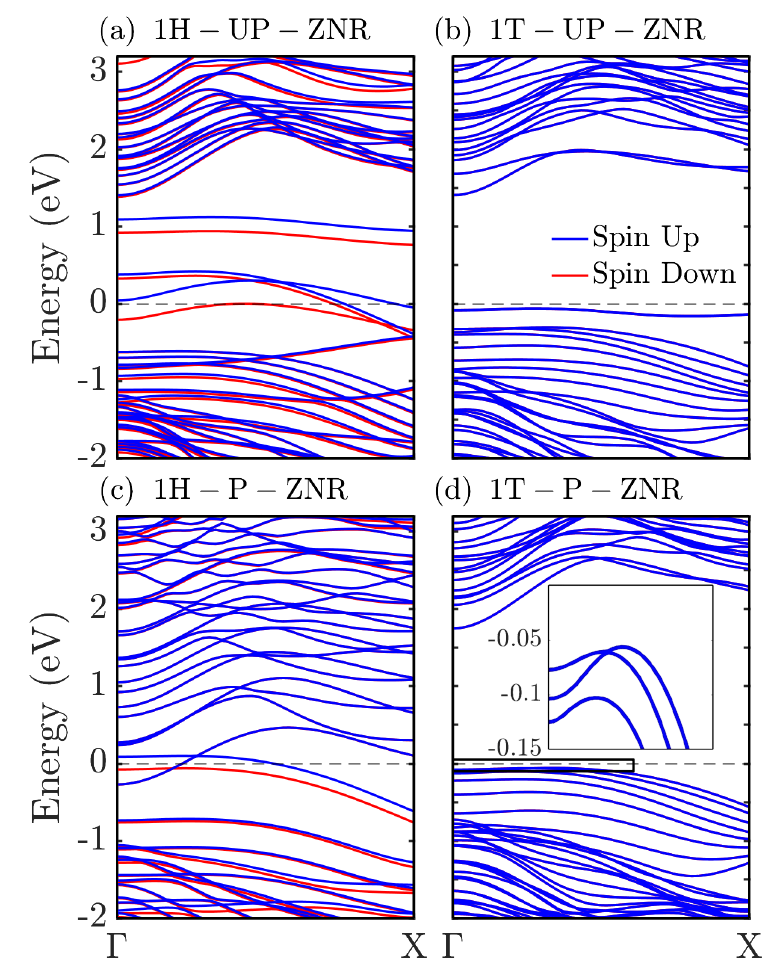}
\caption{Spin-polarized band structures of (a) unpassivated 1H, (b) unpassivated 1T, (c) passivated 1H and (d) passivated 1T $N=8$ GaSe ZNRs.}
\label{fig:supp7}
\end{figure}

\begin{figure}[h!]
\centering
\includegraphics[width=12.2cm]{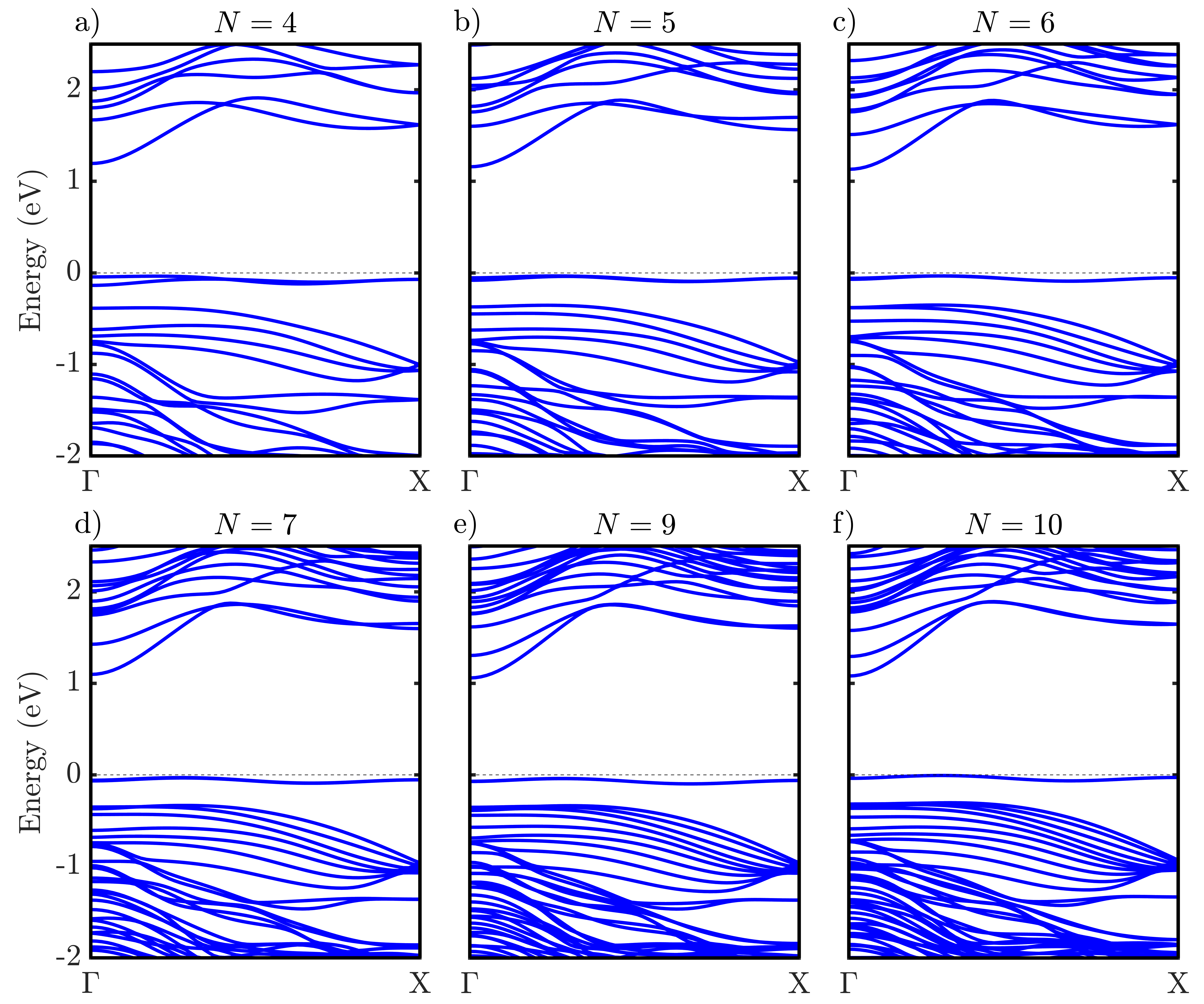}
\caption{Spin-polarized band structures of InSe 1T-UP-ZNRs for  ${N} = 4,5,6,7,9,10$.}
\label{fig:supp8}
\end{figure}

\begin{figure}[h!]
\centering
\includegraphics[width=12.2cm]{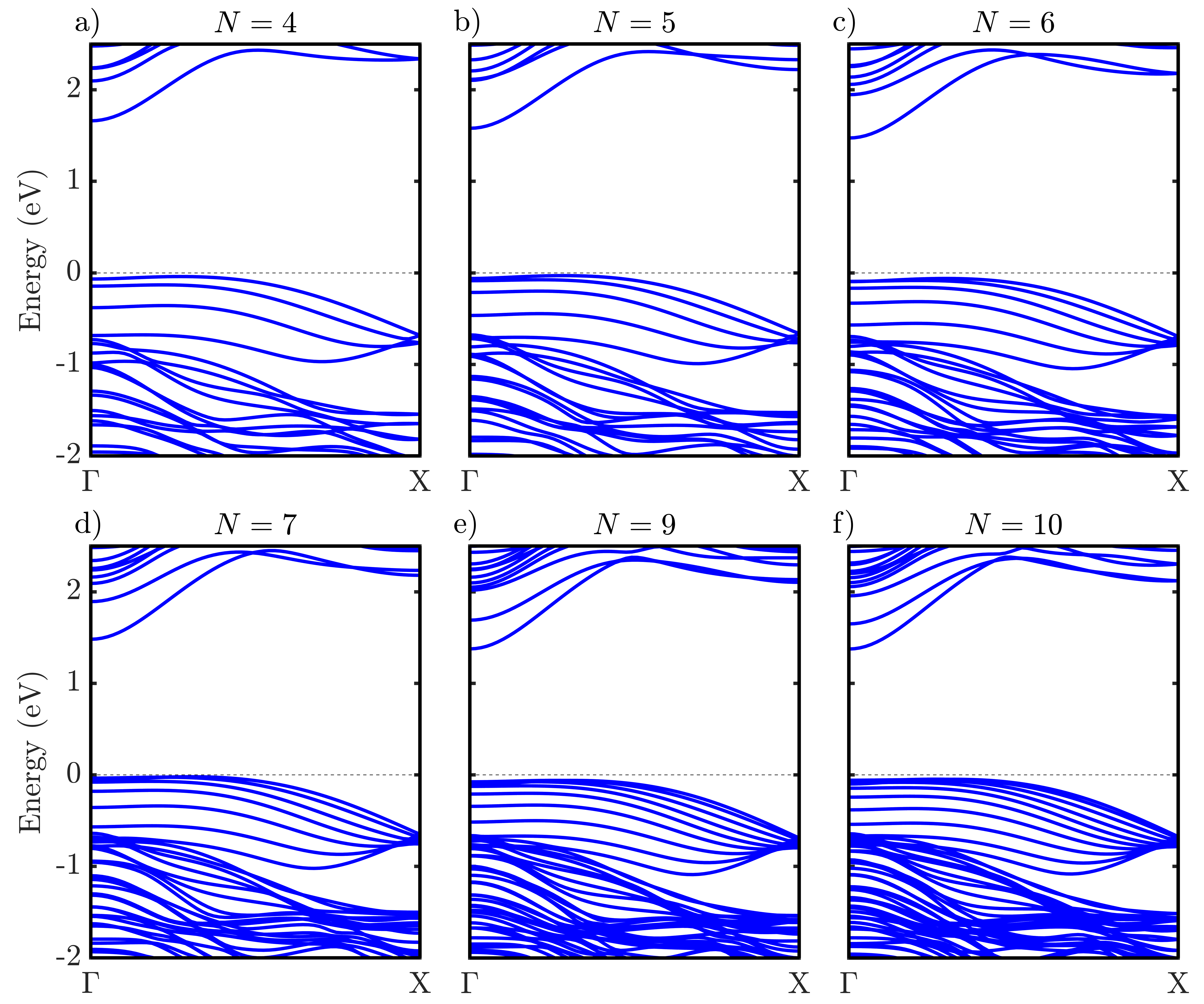}
\caption{Spin-polarized band structures of InSe 1T-P-ZNRs for  ${N} = 4,5,6,7,9,10$.}
\label{fig:supp9}
\end{figure}

\begin{figure}[h!]
\centering
\includegraphics[width=13cm]{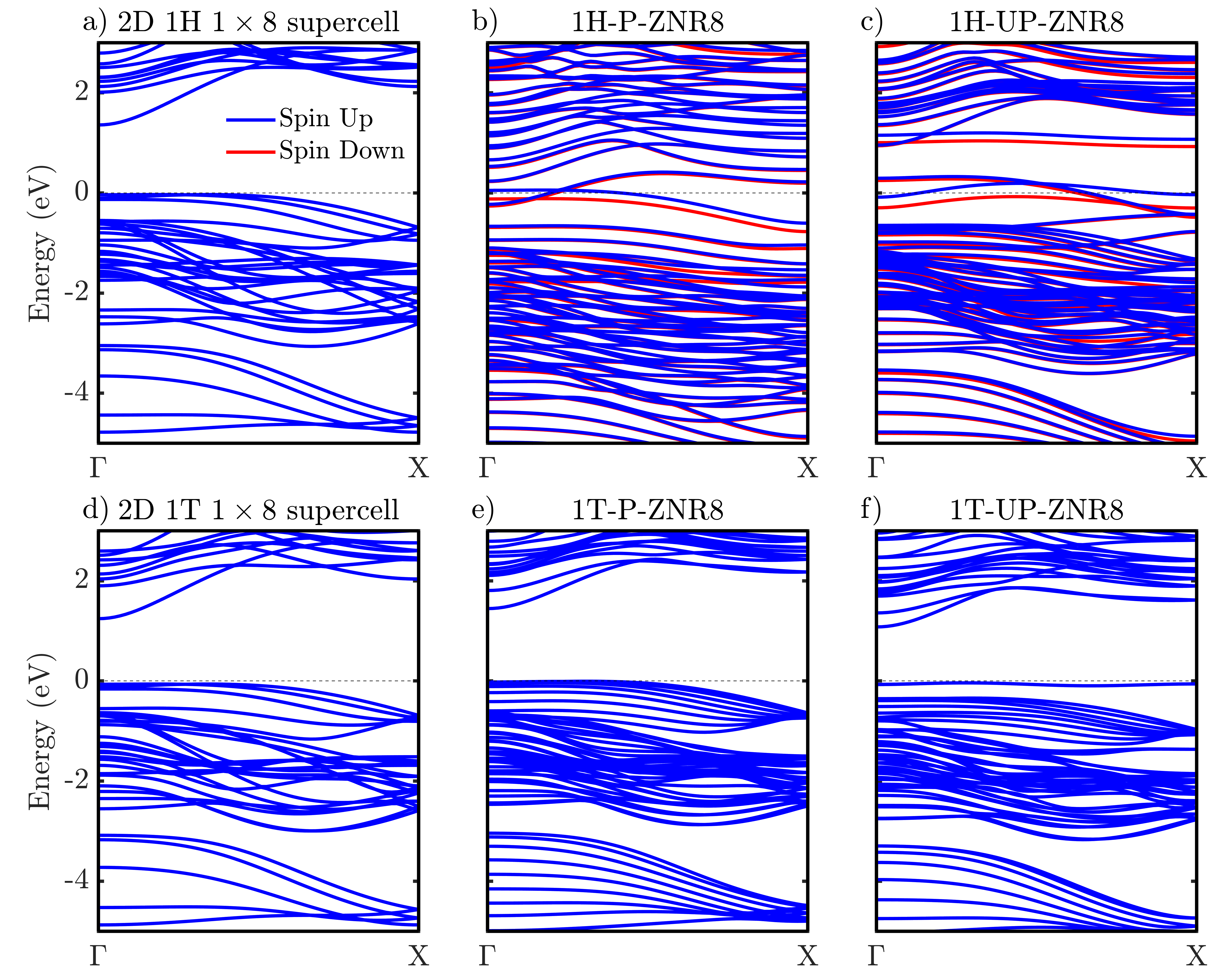}
\caption{Spin-polarized band structures of InSe a) 2D 1H $1 \times 8$ supercell, b) 1H-P-ZNR8, c) 1H-UP-ZNR8, d) 2D 1T $1 \times 8$ supercell, e) 1T-P-ZNR8, and f) 1T-UP-ZNR8.}
\label{fig:supp10}
\end{figure}
\clearpage
\end{document}